\documentclass{aa}
\usepackage[varg]{txfonts}
\usepackage{lscape}
\usepackage{graphicx}
\usepackage{natbib}
\usepackage{multicol}
\usepackage[debugshow]{supertabular}
\graphicspath{{./figures/}}
\defcitealias{Saglia10}{S10}
\defcitealias{Blana16a}{B17}

\begin{document}
\title{
{Evidence for non-axisymmetry in M31\\from\\wide-field kinematics of stars and gas}
\thanks{This paper includes data taken at The McDonald Observatory of The University of Texas at Austin.}\fnmsep 
\thanks{This research was supported by the DFG cluster of excellence 'Origin and Structure of the Universe'.}}

\titlerunning{Hints for a bar in M31 kinematics and morphology}

\author{M. Opitsch \inst{1,2,3}
\and M.H. Fabricius \inst{1,2}
\and R.P. Saglia \inst{1,2}
\and R. Bender \inst{1,2}
\and M. Bla\~na \inst{1}
\and O. Gerhard \inst{1}
}

\institute{Max Planck Institute for Extraterrestrial Physics, Giessenbachstr., D-85748, Garching, Germany
\and Universit\"ats-Sternwarte M\"unchen, Scheinerstr. 1, D-81679, Munich, Germany
\and Excellence Cluster Universe, Boltzmannstr. 2, D-85748, Garching, Germany}


\abstract{} 
{As the nearest large spiral galaxy, M31 provides a
unique opportunity to learn about the structure and evolutionary history of
this galaxy type in great detail. Among the many observing programs aimed at
M31 are microlensing studies, which require good three-dimensional models of
the stellar mass distribution. Possible non-axisymmetric structures like a bar
need to be taken into account. Due to M31's high inclination, the bar is
difficult to detect in photometry alone. Therefore, detailed kinematic
measurements are needed to constrain the possible existence and position of a
bar in M31.}
{We obtained $\approx$ 220 separate fields
with the optical IFU spectrograph VIRUS-W, covering the whole bulge region of
M31 and parts of the disk. We derive stellar line-of-sight velocity
distributions from the stellar absorption lines, as well as velocity
distributions and line fluxes of the emission lines H$\beta$, [OIII] and [NI].
Our data  supersede any previous study in terms of spacial coverage and
spectral resolution.}
{We find several features that are
indicative of a bar in the kinematics of the stars, we see intermediate
plateaus in the velocity and the velocity dispersion, and correlation between
the higher moment $h3$ and the velocity. The gas kinematics is highly
irregular, but is consistent with non-triaxial streaming motions caused by a
bar. The morphology of the gas shows a spiral pattern, with seemingly lower
inclination than the stellar disk. We also look at the ionization mechanisms of
the gas, which happens mostly through shocks and not through starbursts.}{}


\maketitle
\section{Introduction}
\subsection{Recent surveys of M31}
Because of its proximity, M31 is the best case after the Milky Way to learn about the detailed evolutionary history of a large spiral galaxy. Therefore, M31 has been studied by several major surveys in recent years,
combining large-scale photometry with pointed spectroscopic observations. Two of these large programs are the Spectroscopic and Photometric Landscape of Andromeda's Stellar Halo survey  (SPLASH) \citep{Guhathakurta05, Guhathakurta06, Gilbert06} 
and the Pan-Andromeda Archaeological Survey (PAndAS) \citep{McConnachie09}. 
These surveys studied the stellar halo of M31 in great detail, to measure its global properties \citep{Gilbert12, Gilbert14} and look at structures within the halo like 
the Giant Southern Stream \citep{Gilbert09}. This provided information about the formation history of M31.
One of the results of the PAndAS survey was that the dwarf galaxies around M31 all lie in a thin plane \citep{Ibata13}, which poses problems for the 
current understanding of galaxy formation. 
The Panchromatic Hubble Andromeda Treasury (PHAT) program \citep{Dalcanton12} looked at the disk of M31
and measured photometry for 117 million individual stars \citep{Williams14}. 
The Herschel Exploitation of Local Galaxy Andromeda (HELGA) \citep{Fritz12} observed M31
in the far infrared and sub-millimeter wavelengths, measuring the distributions of dust \citep{Smith12} and molecular clouds \citep{Kirk15}.
Microlensing events and variable stars were monitored by the PAndromeda survey \citep{Lee12, Lee14b, Lee14a}. \\
All these surveys help to understand the structure, the accretion history and the history of star formation within M31.\\
M31 is classified as unbarred spiral SA(s)b by \citet{deVaucouleurs91}, however, there has been evidence from photometry and kinematics that it is barred, 
see section \ref{sec:Bar_intro}. In this paper we present new spectroscopic observations of M31. The paper is focused on the description of the data and the hints for a bar that can be seen there. A discussion of other possible explanations, like a superposition of disks and rings, goes beyond the scope of this paper and will be 
presented in future papers based on the dataset described here.

\subsection{Microlensing studies towards M31} 
The current $\Lambda$CDM paradigm \citep{PlanckCollaboration16} has been very successful in explaining both the large scale structure 
of the universe and the observed properties of galaxies. One component of this model is the 
cold dark matter, which resides in halos around galaxies.
The nature of this dark matter is not yet understood. The currently preferred candidates for 
dark matter are non-baryonic elementary particles, so-called WIMPs (Weakly Interacting Massive Particles), see e.g. \citet{Bertone10} and references therein. 
Baryonic candidates for dark matter are mostly ruled out, because the fraction of baryonic matter in the universe is only 15$\%$ of the total matter \citep{PlanckCollaboration14}. 
This baryonic dark matter fraction could be composed of large astrophysical objects, like brown dwarfs, Jupiter-sized planets or black holes, 
collectively known as Massive Astrophysical Compact Halo Objects (MACHOs) \citep{Griest91}.
In order to place firm constraints on the existence of these massive objects, microlensing studies have 
been conducted. In such a microlensing event, a MACHO passes between a bright object and the earth, the light from the source objects is deflected by the gravity of the MACHO, which 
leads to a perceived increase in brightness of the source object. By observing such events, the number density
of MACHOs can be determined \citep{Paczynski86}. The MACHO
survey \citep{Alcock93} found that the contribution of MACHOs to the total halo mass of the Milky Way
is 20$\%$ \citep{Alcock00}, with an average MACHO mass of $\approx$ 0.4 M$_{\odot}$. The EROS and EROS-2 projects \citep{Aubourg93, Afonso03} measured a significantly lower fraction for the same masses, at less than 8$\%$ \citep{Tisserand07}.
The OGLE survey \citep{Udalski92} found a fraction that is comparable to the one measured by the EROS-survey \citep{Wyrzykowski11}, less than 7$\%$ for MACHO masses lower than 1 M$_{\odot}$. 
In addition to these results on the galactic halo, the galactic bulge microlensing surveys within OGLE-III \citep{Wyrzykowski15}, MOA-II \citep{Sumi13} and EROS-2 \citep{Hamadache06} have become major tools for understanding the structure of the Milky Way, see e.g. \citet{Wegg16} for a new analysis. \\
Several microlensing surveys have also been focused towards M31, like the Pan-STARRS 1 survey of Andromeda (PAndromeda) \citep{Lee12} and the
Wendelstein Calar Alto Pixellensing Project (WeCAPP) \citep{Lee15}.
A total of 56 events have been detected in M31 \citep{Lee15}.
These events do not have to be caused by MACHOs, they can also happen due to lensing 
by other stars in M31, so-called self-lensing \citep{Riffeser06}. To get an idea of the amount of events that are caused by self-lensing, a proper
understanding of the three-dimensional distribution of the stars is needed. Models with different mass distributions of the galaxy result in widely varying predictions of the 
event rate of self-lensing and lensing through MACHOs. It is therefore important to 
model the stellar mass distribution in the galaxy as accurately as possible. Based on the data presented in this paper and numerical models from \citet[hereafter B17]{Blana16a} and \citet{Blana16b},
new predictions for microlensing events will be presented in a future paper \citep{Riffeser17}.

\subsection{The bar in M31}
\label{sec:Bar_intro}
A large fraction of disk galaxies in the local universe is barred, ranging from about 50$\%$ in the optical \citep{Barazza08} to about 60$\%$ to 70$\%$ in the infrared \citep{Eskridge00, Menendez-Delmestre07}.
It is now thought that global instabilities in the disk lead to the quick formation of bars. 
In this process, the m=2 mode grows strongly by swing-amplification and forms a long-lasting bar non-linearly \citep{Sellwood93, Sellwood13}.
Over time, the inner part of the bar goes through a buckling phase, which is a short but violent 
vertical instability not long after bar formation \citep{Combes81, Combes90, Raha91, Merritt94, Athanassoula02, Debattista05}.
The instability bends out of the plane of the disk, then settles back 
to the plane, redistributing energy to smaller spatial scales and to higher stellar velocity dispersion, thereby thickening the bar \citep{Raha91}.
The buckled part of the bar becomes a three-dimensional so-called boxy/peanut shaped (B/P) bulge, the part that has not buckled is referred to as the thin or flat bar.
While this buckling phase is frequently seen in simulations, it has only recently been detected in observations by \citet{Erwin16} for two 
local spiral galaxies.

While the Milky Way was originally thought of as unbarred, it is now widely accepted that it
contains a bar. For a review, see \citet{Gerhard15}, and references therein. Recently, signs for a bar have also been detected in the innermost parts 
of the third large spiral galaxy in the Local Group, M33 \citep{Hernandez-Lopez09}.

{The question if M31 also contains a bar or not has not been settled, e.g. \citet{Widrow03} refer to the bulge as ``barlike'', while \citet{Kormendy10} classify M31 as containing a classical bulge.}
{If present, a} bar is not easily detected in images of M31 because of its high inclination of 77$\degr$ \citep{Walterbos87}. 
This is too high to see a bar directly in the image, but too low to recognize its 
shape above and below the stellar disk, as is possible in an edge-on view \citep{Athanassoula06}. Nevertheless, the boxy appearance of the 
isophotes in near-infrared images already is a hint at the existence of a bar \citep{Beaton07}. However, boxy isophotes do not need to be caused by bars, there are numerous examples of early-type galaxies 
that are boxy without having a bar, see e.g. \citet{Kormendy09}.\\
It is possible to detect a bar in a galaxy with 
an inclination similar to the one of M31, \citet{KuziodeNaray09} investigated the galaxy NGC 2683 with a similar inclination to M31, by studying ionized gas velocities and the overall morphology. By combining the photometric and kinematic data, they 
found evidence for the presence of a bar in NGC 2683 and constrained its orientation and strength.\\
According to \citet{Stark94}, there are three arguments for a {triaxial structure} in M31:
\begin{enumerate}
 \item There is a twist in the inner isophotes in the bulge with respect to the outer disk, first seen by \citet{Lindblad56}. He was subsequently the first one to claim that M31 has a bar. These twists cannot be 
 reproduced by a rotationally symmetric distribution of stars \citep{Stark77}.
 \item The velocities of the HI gas are not symmetric about the minor axis \citep{Rubin71}.
 \item The ionized gas has the appearance of a spiral pattern, which is rounder than 
 the appearance of the disk, as seen by \citet{Jacoby85}, \citet{Boulesteix87} and \citet{Ciardullo88}.
 \end{enumerate}
\citet{Stark77} showed that the features measured by \citet{Lindblad56} can be 
explained by a family of {triaxial bulge} models. 
\citet{Stark94} narrowed down these models by simulating the velocities of the gas 
in this potential.\\
\citet{Berman01} {and \citet{Berman02}} simulated the gas velocities in the {triaxial bulge} potential that was derived using the method of \citet{Stark77} and they are in agreement with 
the non-circular gas velocities in the inner disk.
{Their model has a fast pattern speed of 53.7 km s$^{-1}$ kpc$^{-1}$. Therefore, it would 
be more fitting to call their triaxial bulge a bar. In our understanding, a triaxial bulge would be a non-rotating structure.
Most observed bars have after the buckling phase a three-dimensional inner part of the bar, the B/P-bulge, and a flat outer part (see e.g. \citet{Athanassoula05} or \citet{Martinez-Valpuesta06}). 
In this nomenclature, which we adopt throughout this paper, the ``triaxial bulge'' of the models by \citet{Berman01} and 
\citet{Berman02} is a B/P-bulge.}
According to \citet{Gordon06}, the {model by 
\citet{Berman01}} explains the morphology of dust in M31,
with spiral arms emerging from the bar. However, the fact that the two prominent dust rings do not share the same center, which also does not coincide with the optical center of M31, led \citet{Block06}
to propose a different scenario, where these rings were not created by a bar, but instead are shock waves due to the collision of the small companion galaxy M32 with M31.

\citet{Athanassoula06} tested four different bar models and qualitatively compared the velocities to HI kinematics from \citet{Rubin70}, \citet{Brinks84a} and \citet{Brinks84b}, and the overall morphology to observations in the near infrared by \citet{Beaton07}.
They found that in order to explain the boxy appearance of the 
isophotes in \citet{Beaton07}, a classical bulge needs to be present.
The {triaxial bulge} 
seen by \citet{Lindblad56}, \citet{Stark77} and \citet{Stark94}, corresponds to the B/P bulge from \citet{Athanassoula06}. 
The fact that the boxy isophotes in \citet{Beaton07} do not coincide with the disk argues for a misalignment of the 
bar and disk.\\ 
While the arguments for a bar in \citet{Athanassoula06} are mostly qualitative, a more quantitative result was obtained by \citetalias{Blana16a}, who tested 84 different models and compared them to 
3.6$\mu$m infrared photometry from \citet{Barmby06}, HI kinematics from \citet{Chemin09} and \citet{Corbelli10}, as well as
stellar kinematics from \citet[hereafter S10]{Saglia10} and data from the work presented in this paper. 
Again, they rule out solutions which do not contain a classical bulge component embedded within the B/P bulge, finding for the mass of the classical bulge
$M_{class}=$1.0 - 1.4 $\cdot$ 10$^{10} M_{\odot}$ and for the half mass radius $r_{class} = 0.5 - 1.1$ kpc. In the preferred model in \citetalias{Blana16a}, 
the mass of the classical bulge component is $M_{class, best}$ = 1.1 $\cdot10^{10}$ M$_{\odot}$ and the half-mass radius is $r_{class, best}$= 0.53 kpc,
for the B/P component, the parameters are $M_{B/P, best}=2.2 \cdot 10^{10} M_{\odot}$ and $r_{B/P, best}=1.3 $ kpc. 
The projected position angle of the bar is $PA_{bar}=55.7 \degr$, which is 17.7$\degr$ more than the disk position angle of PA=38$\degr$ \citep{deVaucouleurs58}.
The bar in the model by \citetalias{Blana16a} started buckling about 2 Gyr after it formed, the buckling phase ended about 1 Gyr later. This happened at least 0.8 Gyr ago, but it could have happened earlier, because once the buckling 
phase is over, the galaxy evolves very slowly with time, so it can't be said when these events happened exactly.
The intrinsic length of the bar is 1000\arcsec, which projected onto the sky with M31's orientation and inclination becomes 600\arcsec.\\
According to the models by \citet{Athanassoula06} and \citetalias{Blana16a}, there are at least the following separate structural stellar components in M31, from the innermost to the outermost:
\begin{enumerate}
 \item A classical spherical bulge in the center,
 \item a B/P bulge, which is the inner thicker part of the bar,
 \item a thin bar, this is the outer part of the bar, 
 \item a disk and 
 \item a halo. 
\end{enumerate}

This is the first paper in a series of papers about our observations of M31, this one covers the description of the data, the kinematics, and the gas fluxes, in an accompanying paper we will present results on the stellar populations. The data will also be made available online. \\
This paper is structured as follows:
In section \ref{sec:Observations}, our observations of M31 are described, before we present the methods used to fit the kinematics in section \ref{sec:Methods}. 
Section \ref{sec:StellarKinematics} then presents the results for the stellar and gas kinematics, as well as the gas morphology. In section \ref{sec:Bar}, we search for arguments for the bar in the data, before 
summarizing our findings in section \ref{sec:Conclusions}.\\
We adopt a distance to M31 of 0.78 $\pm$ 0.04 Mpc \citep{deGrijs14}, an inclination of 77$\degr$ \citep{Walterbos87} and 
a heliocentric velocity of -300 $\pm$ 4 km s$^{-1}$  \citep{deVaucouleurs91}. 

\section{Observations}
\label{sec:Observations}
\subsection{The IFU spectrograph \texttt{VIRUS-W}}
The research of this project was carried out with the IFU spectrograph \texttt{VIRUS-W} \citep{Fabricius12} mounted on the 2.7m telescope at the 
McDonald observatory. The integral-field unit consists of
267 fibers which are arranged in a rectangular hexagonal dense-pack scheme
with a filling factor of 1/3. The field-of-view of the instrument is 105\arcsec x 55\arcsec at the 2.7m telescope, with the long edge of the fiberhead aligned along the east-west axis. Each 
fiber covers a circle with diameter 3.2\arcsec on sky. The actual spectrograph has two different resolution modes, each realized with a Volume Phase
Holographic (VPH) grating. We use the high-resolution mode, where the grating has a line frequency of 3300 lines per millimeter and a resolution of R $\approx$ 9000, which corresponds to an instrumental dispersion of $\sigma_{inst}=15$ km s$^{-1}$.
For a VPH a change of the grating angle results in an effective change of the blaze function, such that the throughput for a specific wavelength range is optimized. For our observations, we adjusted the grating angle to 353$\degr$ after some testing to get moderately high throughput at the wavelength of $H\beta$ at 4861 \AA, with the maximum of the throughput being 
between 4900 \AA \ and 5100 \AA.
The complete wavelength range is 4802 \AA \ to 5470 \AA. In this range, we see the emission lines H$\beta$ at 4861 \AA, the doublet of forbidden 
lines of doubly ionized oxygen [OIII]$\lambda \lambda$4959, 5007 and the doublet of the forbidden nitrogen lines [NI]$\lambda \lambda$ 5198, 5200.

\subsection{Description of the observations}
\begin{figure}
\resizebox{\hsize}{!}{\includegraphics{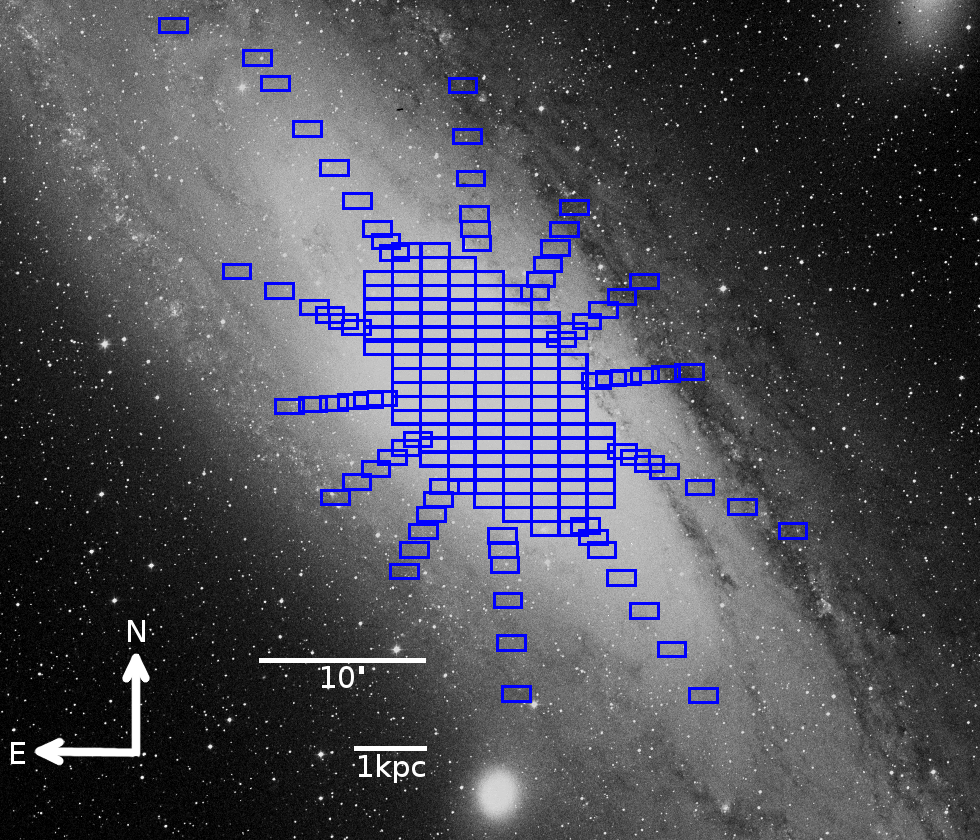}}
\caption[Observed pointings]{Observed pointings of M31 overlaid on a Digitized Sky Survey\footnotemark \ V-band image.}
\label{fig:total_observations}
\end{figure} 
\footnotetext{The image has been taken from \url{http://archive.stsci.edu/cgi-bin/dss_form}. The Digitized Sky Surveys were produced at the Space Telescope Science Institute under U.S. Government grant NAG W-2166. The images of these surveys are based on photographic data obtained using the Oschin Schmidt Telescope on Palomar Mountain and the UK Schmidt Telescope. The plates were processed into the present compressed digital form with the permission of these institutions. \textcopyright 1995 by the Association of Universities for Research in Astronomy, Inc.}
We observed 198 pointings in four separate observing runs, in October 2011, October 2012, February 2013 and August 2013. The positions of all observed pointings are shown in Fig. \ref{fig:total_observations}. \\ 
The observations consist of a completely covered area and six stripes extending further out. The angles of these stripes are $35\degr$ (approximately 
the disk major axis), $65\degr$, $95\degr$, $125\degr$ (approximately the disk minor axis), $155\degr$ and $185\degr$. The completely covered region corresponds to the area where the bulge dominates the overall light \citep{Kormendy99}. Therefore, all pointings in the completely covered area will be called ``bulge pointings'' and 
the ones in the stripes ``disk pointings''. 
Along the major axis, we reached approximately one disk scalelength of $r_d=24'=5.3$ kpc \citep{Courteau11}. We did not dither our observations, because we were primarily interested in covering a large area of M31 efficiently. 
We observed each galaxy pointing with an exposure time of 10 minutes, except for the central pointing, which was only observed 
for 5 minutes, because it covers the bright nucleus of M31, where sufficient signal-to-noise values were already reached with this shorter exposure time. Before and after each galaxy pointing, we 
nodded the telescope away from the galaxy to a sky position, which was exposed for 5 minutes. The seeing varied between 1.3\arcsec and 3.0\arcsec during the observations. At the beginning 
and at the end of each observation night, we took bias, flat and arc images.\\

\subsection{Data reduction}
\label{sec:Reduction}
The data reduction follows the standard procedure for \texttt{VIRUS-W} as described in \citet{Fabricius14}.
It uses the \texttt{fitstools} package \citep{Gossl02} and the \texttt{Cure} pipeline developed for \texttt{HETDEX} \citep{Hill04}. \\
First, master biases, flats and arcs are created by taking the mean of the individual images for each morning and evening.
The master bias frames are subtracted from all other frames. \\
\texttt{Cure} traces the fiber positions on the master flat frames and then extracts the positions of the spectral line peaks along these traces in the 
master arc frames. 
To model the distortion and the spectral dispersion, a two-dimensional seventh
degree Chebyshev polynomial is used. The resulting model transforms between pixel 
positions on the detector and fiber-wavelength pairs and vice-versa.\\
Foreground stars, which appear as brightly illuminated fibers, are clipped using a $\kappa$-$\sigma$-clipping 
method.\\
For the wavelength calibration, 27 lines are used, they are listed in Table \ref{tab:HgNe} in the Appendix.\\
Having traced the fiber positions and calibrated the wavelengths, the spectra are now 
extracted from the science frames by walking along the trace positions and averaging the 
values in a 7 pixel wide aperture. The extraction is performed in $\ln(\lambda)$-space, the 
step width corresponds to 10 km s$^{-1}$. \\
The flatfield frames are extracted in the same way as the data frames.\\ 
The fiber to fiber throughput variation and the vignetting are corrected by dividing each observation by 
the corresponding flatfield observation. The flatfields are very stable over the course of one observing night, with the deviations being on the order of 0.03$\%$.
However, the resulting flat-fielded spectra still exhibit the rather strong variation of sensitivity as function of wavelength that is due to
the strongly peaked diffraction efficiency of the VPH grating. This would complicate the later throughput calibration. Therefore, in the next step
the spectra are divided at each wavelength by the mean value of the flat field spectrum at that wavelength, where the mean is taken across all fibers in the flatfield observation. 

To take care of the sky, we pointed the telescope to a sky position before and after each galaxy pointing. The sky frames are fiber-extracted and flat-fielded in the same way as the galaxy observations. They are quite homogeneous after being flat-fielded, with a root-mean-square of the order of 1$\%$. We average the two sky pointings observed before and after each galaxy pointing into one. 
The difference between the two sky pointings is very small, about $3\%$. We then subtract this flat-fielded sky image 
from the flat-fielded galaxy image.

Because the different observing runs took place in different months of the year, we also have to correct during the extraction for the relative motion of the earth around the sun.
We use the web-tool by Edward Murphy\footnotemark \ based on an algorithm described in \citet{Meeks76} to calculate the relative velocity of the Earth towards M31 at the 
time of the observation. For each observing run, we use the value for the mean date of the run. We calculate the correction relative to the run in October 2011, because for that 
run, the correction is $c_{Oct11}=0$ km s$^{-1}$. The correction for the run in October 2012 is $c_{Oct12}=-3.6$ km s$^{-1}$, for the run in February 2013 it is $c_{Feb13}=19.1$ km s$^{-1}$  and for the run in August 2013 it is 
$c_{Aug13}=-28.1$ km s$^{-1}$. \footnotetext{\url{https://www.astro.virginia.edu/~emm8x/utils/vlsr.html}}\\
The sky position for every fiber in each pointing of M31 is determined following \citet{Adams11}.
The accuracy of this method was estimated by \citet{Adams11} to be 0.21\arcsec, much less than the 
fiber diameter of \texttt{VIRUS-W} of 3.2\arcsec.
The coordinates are converted to distance in arcseconds relative to RA=10:41:07.04 and DEC=41:16:09.41. This is the coordinate of the pointing covering the center of M31 in our observations. \\ 
In the central region, the signal-to-noise ratio (S/N) of each fiber spectrum lies well above 30. This remains true out to a radius of about 140\arcsec \ along the major axis and 100\arcsec \ along the minor axis. 
At larger radii m fiber spectra are then spatially  binned to reach a minimum S/N of 30 using the Voronoi-binning method by \citet{Cappellari03}, resulting in a total of 7563 binned spectra. 

\section{Methods}
\label{sec:Methods}
\subsection{Measuring the stellar kinematics with \texttt{pPXF}}
The kinematics is measured using the routines \texttt{pPXF} (penalized PiXel Fitting) by \citet{Cappellari04} and \texttt{GANDALF} (Gas AND Absorption Line Fitting) by \citet{Sarzi06}.
\texttt{GANDALF} uses \texttt{pPXF} as its first step.
\texttt{pPXF} measures the stellar kinematics by broadening a weighted sum of template star spectra with a trial line-of-sight velocity distribution (LOSVD) and subsequently changing the parameters of the 
LOSVD until the residuals between the measured and the model spectrum are minimized. 
We use spectra from 41 kinematic standard stars obtained with \texttt{VIRUS-W}. They are listed in Table \ref{tab:kinematic_standards} in the Appendix. The information about the stars is taken from the \texttt{ELODIE} \citep{Prugniel07} and \texttt{LICK} \citep{Worthey94} catalogs. 
The coordinates come either from the \texttt{ELODIE} catalog or \citet{vanLeeuwen07}, using the \texttt{SIMBAD} interface \citep{Wenger00}. \\
In \texttt{PPXF}, the LOSVD is expanded as a Gauss-Hermite series following \citet{vanderMarel93} and \citet{Gerhard93}:
\begin{equation}
 \mathcal{L}(v)= \frac{\exp\left(-\frac{(v-<v>)^2}{2 \sigma^2}\right)}{\sigma \sqrt{2 \pi}} \left[1+ \sum_{m=3}^{M} hm \cdot H_m\left(\frac{v-<v>}{\sigma}\right) \right] 
 \label{eq:LOSVD}
 \end{equation}
$H_m$ are the Hermite polynomials and $hm$ the Gauss-Hermite coefficients, the sum is broken off after $M$ entries. We only look at the Gauss-Hermite moments $h3$ and $h4$. \\

\subsection{Fitting the emission lines with \texttt{GANDALF}}
\begin{figure}
\resizebox{\hsize}{!}{\includegraphics{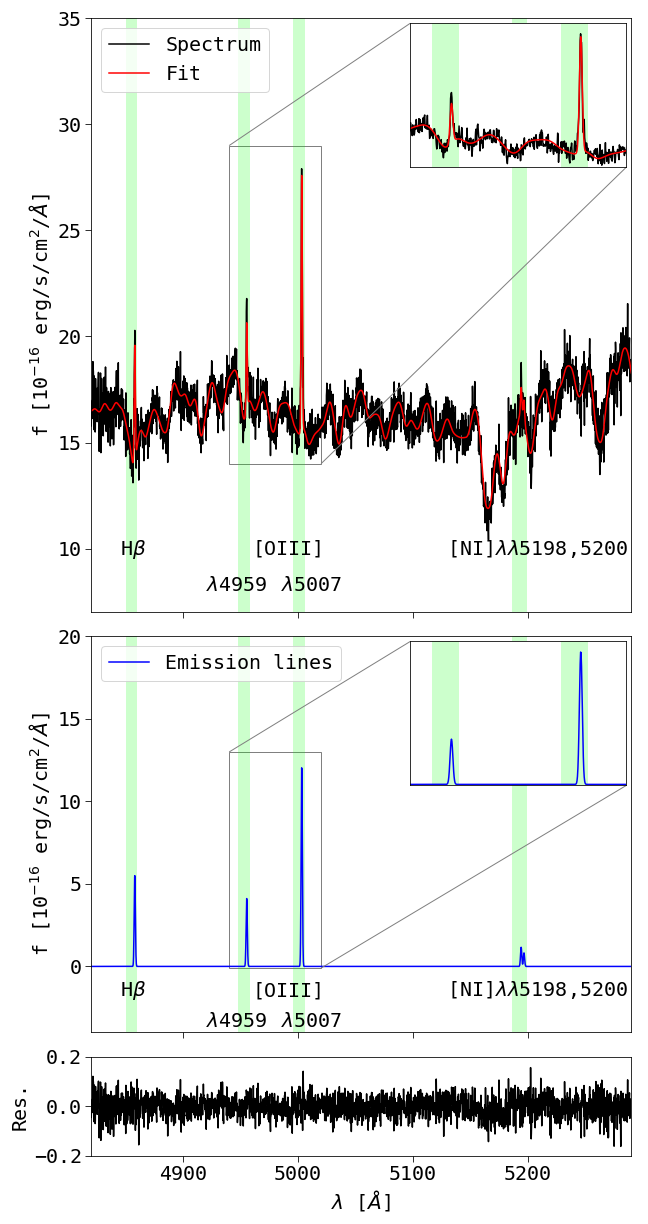}}
 \caption[Example spectrum with \texttt{GANDALF} fit]{A spectrum from the bulge region with the 
 fit by \texttt{GANDALF}. \\ Top: Flux-corrected spectrum (black), best fit  
 by \texttt{GANDALF} (red), which is the sum of the model stellar and the emission line spectra. The green shaded areas are the regions where the emission lines 
 are expected. A zoom into the region of the [OIII] doublet is shown on the right.\\ Middle: The best fit emission 
 lines with a zoom into the region of the [OIII] doublet.\\
 Bottom: The residuals (f$_{measured}$ - f$_{fitted}$)/f$_{measured}$.}
\label{fig:Fit_example}
\end{figure}

The kinematics of the ionized emission lines are fitted with \texttt{GANDALF}
\citep{Sarzi06}. It fits the kinematics of the [OIII]$\lambda$5007 line and ties the kinematics of the other 
emission lines to that line. 
In Fig. \ref{fig:Fit_example}, a fit with \texttt{GANDALF} to a spectrum is plotted.

\begin{figure}
\resizebox{\hsize}{!}{\includegraphics{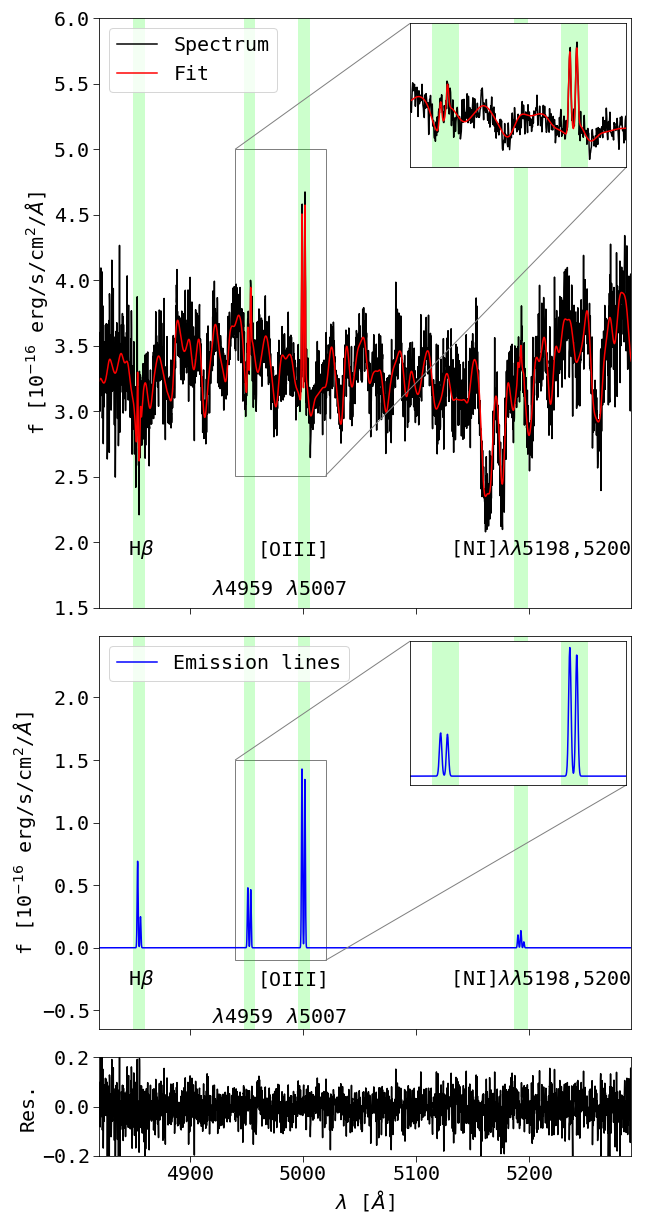}}
\caption[Example spectrum with \texttt{GANDALF} fit with two lines]{A spectrum from the outer edges of the bulge region where the emission lines are split into two lines with almost 
 equal amplitudes. The plot is analogous to Fig. \ref{fig:Fit_example}.}
 \label{fig:Fit_example_two_lines}
\end{figure}

Close examination of the spectra shows that in many regions multiple gas components at different 
radial velocities exist. 
In order to properly treat these multiple peaks, we add another [OIII] component to be fitted with \texttt{GANDALF}.
We also add
a second $H\beta$ component and a second [NI] component. The initial guesses for the gas velocities have to be slightly different for the 
two components, otherwise \texttt{GANDALF} does not fit separate components. An example of a fit with two lines is shown in Fig. \ref{fig:Fit_example_two_lines}. Here, the two lines 
have almost the same amplitude and are clearly separated.

This is not always seen so clearly, there are also cases where one line is stronger than the other or where the two lines are almost blended together or where there is only one line clearly visible with a 
skewed line shape, that can, however, be described by the combination of two lines.

\begin{figure}
 \resizebox{\hsize}{!}{\includegraphics{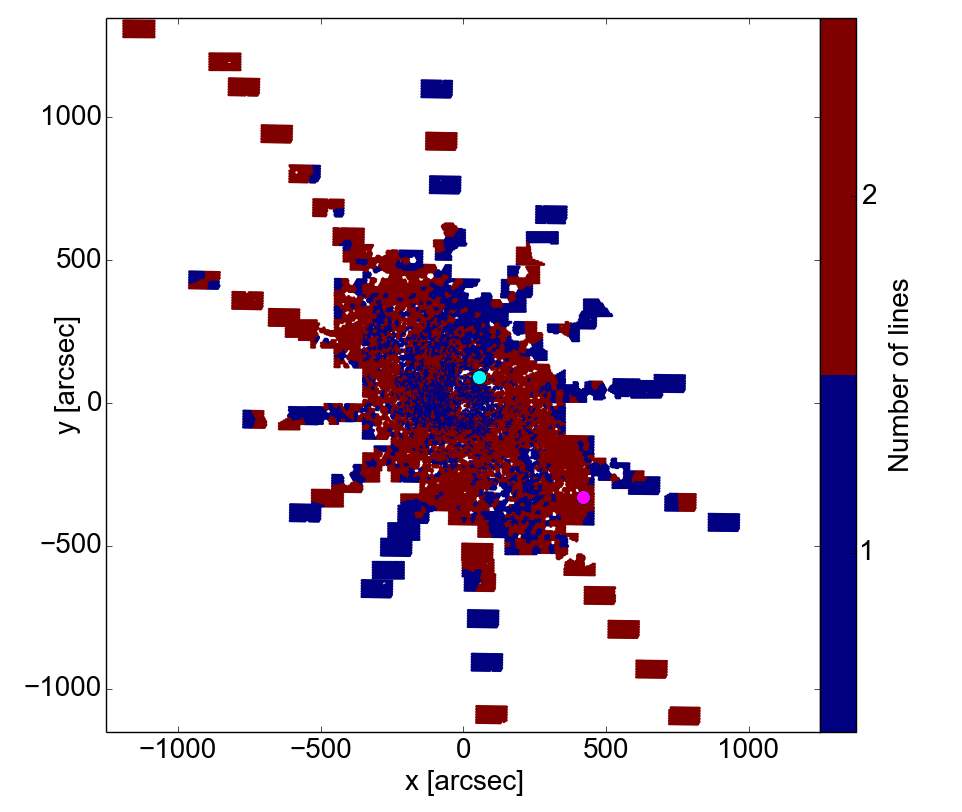}}
 \caption[Regions with one and two lines]{Map of where the emission lines exhibit two peaks (red) and where they only have one (blue). The location of the spectrum with one peak from Fig. \ref{fig:Fit_example} is marked with the cyan dot, 
 the location of the spectrum with two peaks from Fig. \ref{fig:Fit_example_two_lines} is marked with the magenta dot.}
 \label{fig:nlines}
\end{figure}

In Fig. \ref{fig:nlines}, we show where we find one line and two lines. We see only one line in about 3000 bins, while we see two lines in about 3500. 
 
In order to reliably fit the different components, we have to feed
\texttt{GANDALF} information whether only one component is present or if it is
really two.  In order to get these initial guesses, we apply the following
method: First, we cross-correlate a model spectrum only consisting of the
[OIII]$\lambda$5007 line with each spectrum. The program fits the resulting
cross-correlation function with a set of gaussians. These gaussians all have
the same dispersion of $\sigma=20$ km s$^{-1}$  and their peak velocities are
40\,km s$^{-1}$  apart. The program changes the amplitudes of the individual
Gaussians to get the best approximation of the input cross correlated spectrum.
We tell the program to only pick the Gaussian with the largest amplitude to
have an estimate for the one-component fit and the one with the largest
amplitude plus the one with the second largest amplitude for the two component
estimate.  If the amplitude of the second component is less than 0.25 times the
amplitude of the first one, we decide to take the initial guess with only one
component. We also discard unrealistically high velocity dispersions of
$\sigma_{gas} >$ 100\,km s$^{-1}$. We use the central velocities of the
Gaussians as the initial guesses for \texttt{GANDALF}, letting it fit one line
for the cases where we have found only one line and letting it fit two lines
where we have found two lines. \\ After a first iteration, we check all fits
manually, update the initial guesses for the spectra where the fit failed and
let \texttt{GANDALF} fit a second iteration.  This second iteration results in
85 $\%$ of the spectra being fitted correctly, the rest is left out of the
analysis. 

In order to check if the spectra could be contaminated by a contribution of PNe, we looked 
at the catalogue of PNe positions from \citet{Merrett06}. If the position of a PNe is closer than 1.6\arcsec \ (the 
radius of a fiber), we see this bin as affected. Overall, 166 bins are affected, which is 2.5$\%$. All of these spectra show [OIII] lines, about 2/3 of them show two lines, while 1/3 show only one line. The spectra do not look systematically different from the unaffected spectra, there is for example no exceptionally bright line in any of them. We therefore conclude that 
contamination by PNe is negligible.
\subsection{Flux calibration}
\label{sec:flux_calibration}
\begin{figure}
\resizebox{\hsize}{!}{\includegraphics{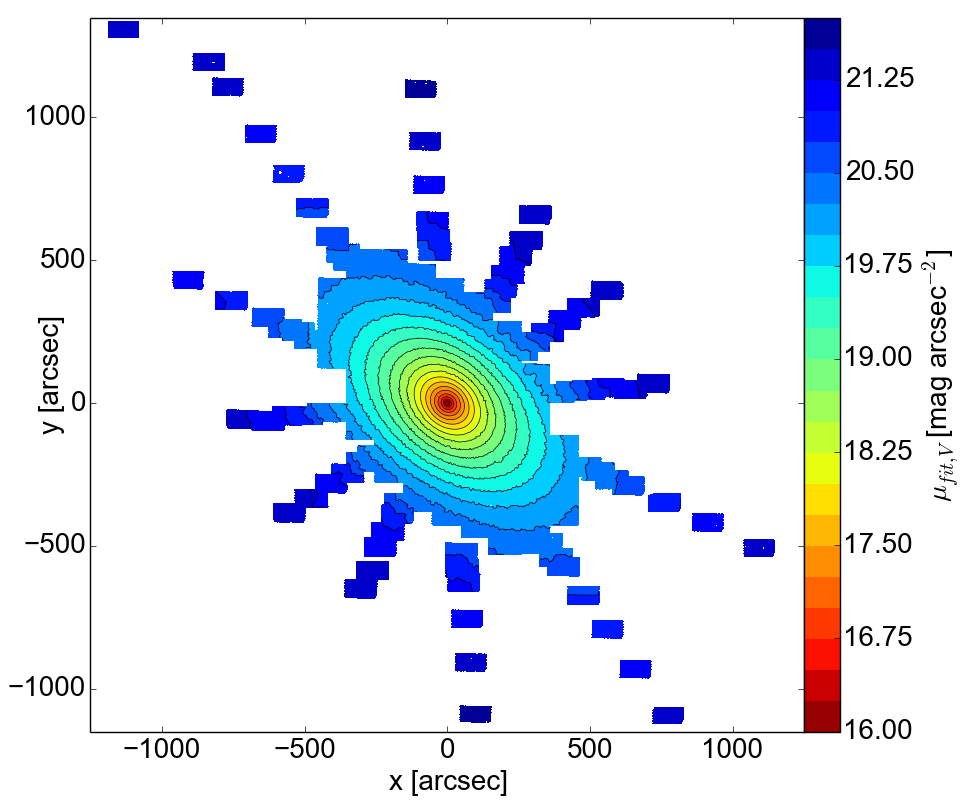}}
\caption[Model image]{Model image of M31, constructed by combining the V-band magnitude from the decomposition of \citep{Kormendy99}
with an ellipse fit to a K-band image by \citetalias{Saglia10}.}
\label{fig:model_image}
\end{figure}

In each observation night, we observe photometric standard stars, they are listed in the appendix in 
Table \ref{tab:photometric_standards}. These spectra are compared to photometrically calibrated 
spectra from the literature \citep{Oke90, LeBorgne03} to get the throughput for the particular observation 
night as a function of wavelength. The throughput curves are calculated with a program from \citep{Mueller14}, the error for 
the throughput curve is about 4$\%$. \\
The atmospheric extinction is also corrected for each individual 
observation.\\
Differences in observing conditions between the individual pointings are taken into account by 
comparing the integrated flux $f_{tot}$ in one spectrum to the flux $f_V$ of a photometric model image. This image is 
constructed by combining the bulge-disk decomposition by \citet{Kormendy99} with an ellipse fit 
to a K-band image performed by \citetalias{Saglia10}. The model image is shown in Fig. \ref{fig:model_image}.

\texttt{GANDALF} calculates 1-$\sigma$ errors in the routine, which are consistent with what we obtain from Monte-Carlo simulations of 
fitting a noiseless spectrum with random added noise. These 1-$\sigma$ errors are the ones tabulated in Table 
\ref{tab:Photometric_gas_data} in the appendix.

\section{Results}
\label{sec:StellarKinematics}
\subsection{Stellar Kinematics}
\begin{figure}
\resizebox{\hsize}{!}{\includegraphics{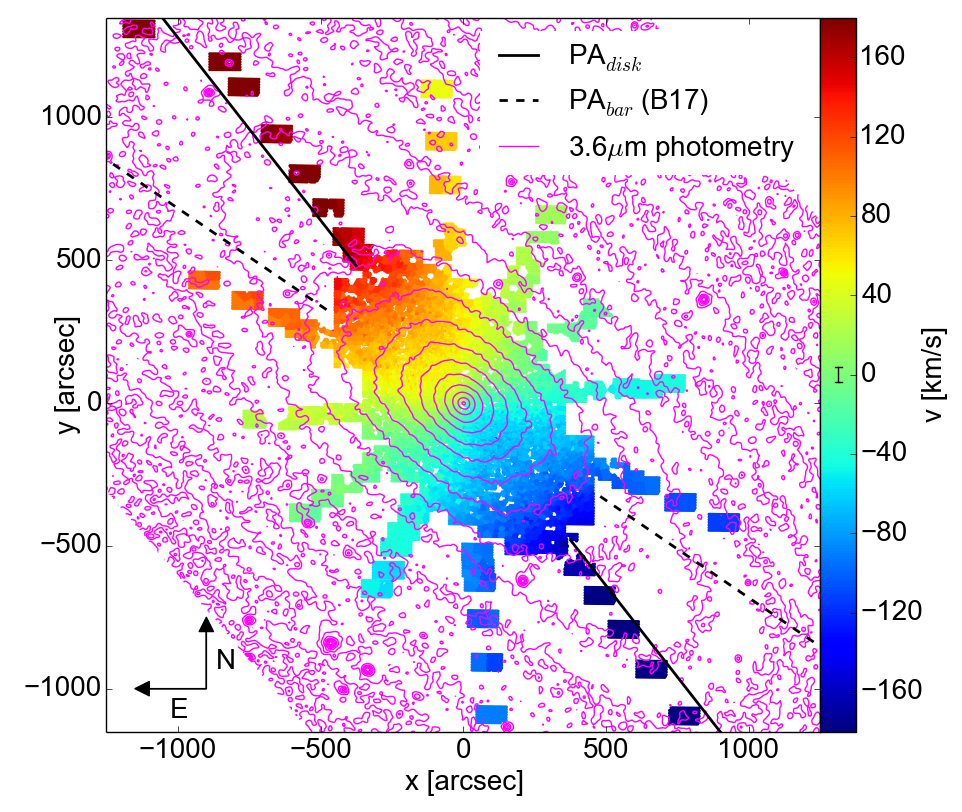}}
 \caption[Stellar velocity]{Map of the stellar velocity corrected for the systemic velocity of M31. The solid straight line is the 
 disk major axis at PA$_{disk}$=38$\degr$, the dashed straight line is the bar major axis at PA$_{bar}$=55.7$\degr$ from \citetalias{Blana16a}. The magenta contours are the surface brightness 
 of the \texttt{IRAC} 3.6$\mu$m image by \citet{Barmby06}. The line in the colorbar is the median errorbar of the velocities.}
 \label{fig:v_star}
\end{figure}

In this section, the stellar kinematics measured with \texttt{GANDALF} is presented. The data will be made available online, an example table is given in Table \ref{tab:data} in the Appendix.
We compare our data with the measurements by \citetalias{Saglia10} in the appendix in section \ref{sec:Saglia}. In general, both datasets agree very well. \\
The stellar velocity map is plotted in Fig. \ref{fig:v_star}, the heliocentric velocity of M31 ($v_{sys}=-300$ km/s; \citealp{deVaucouleurs91}) has been 
subtracted.

\begin{figure}
\resizebox{\hsize}{!}{\includegraphics{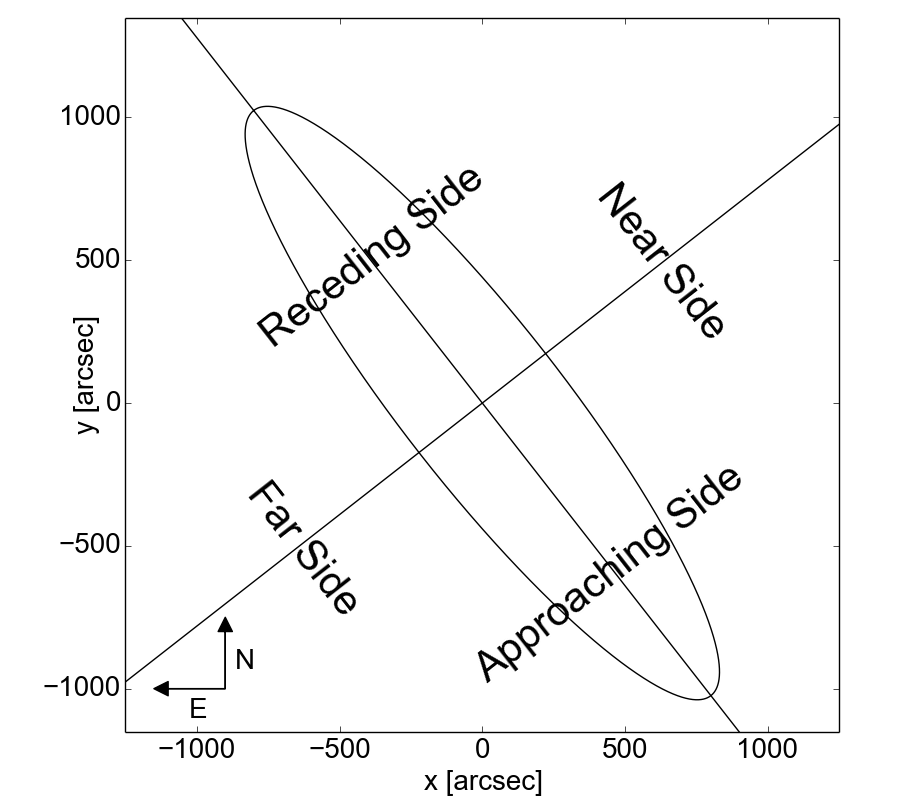}}
\caption[Schematic view of M31]{Schematic view of M31. A generic ellipse with ellipticity 
$\epsilon=0.78$ is plotted, this corresponds to an inclination angle of $i=77$, the disk major axis has position angle $PA=38\degr$. The half of the galaxy to the north-west of the major axis is the near side \citep{Henderson79}.}
\label{fig:M31_schematic}
\end{figure}
\begin{figure}
 \resizebox{\hsize}{!}{\includegraphics{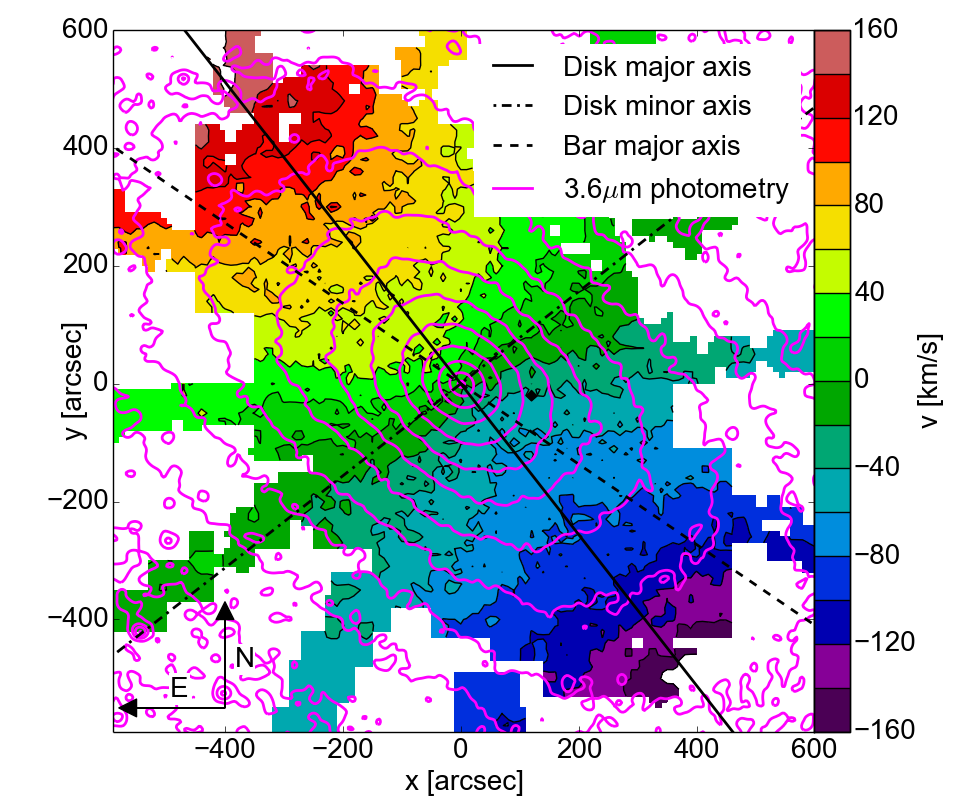}}
 \caption[Stellar velocity with contours]{The map of the stellar velocity from Fig. \ref{fig:v_star} plotted with contours. The solid line is the disk major axis (38$\degr$), the dash-dotted line the disk minor axis (128$\degr$) and the dashed line 
 is the bar major axis (55.7$\degr$). The magenta contours are from the \texttt{IRAC} 3.6$\mu$m image. In this visualisation, the line of zero velocity shows a slight twist in the eastern half of the bulge.}
\label{fig:v_star_log}
\end{figure}

A schematic view of M31 based on \citet{Henderson79} is plotted in Fig. \ref{fig:M31_schematic}, with the naming of the 
receding and approaching sides of M31 taken from the stellar velocity map in 
Fig. \ref{fig:v_star}.
Overall, the stellar velocity field is regular, rotation is clearly visible. 
The velocities increase strongest along the disk major axis, with the highest velocity
in the bulge region being v$_{bulge, max}$ = 136 $\pm$ 4 km s$^{-1}$  in the outermost bulge pointing in the receding side  and the lowest being v$_{bulge, min}$ = -157 $\pm$ 4 km s$^{-1}$  on the opposite side.
The ``bulge region'' is the region of M31 where the bulge-to-total ratio of M31 is higher than 0.5, taken from a decomposition by \citet{Kormendy99}.
The absolute maximum of the velocities is reached with v$_{max}$ = 208 $\pm$ 3 km s$^{-1}$  in the outermost disk major axis disk pointing in the receding side, and the lowest value of v$_{min}$ = -193 $\pm$ 2 km s$^{-1}$  already reached 
in the disk pointing at a radius of 930\arcsec on the approaching side. On that side, the velocity roughly remains at that value to the outermost radii. The median velocity error is d$v=3.8$ km s$^{-1}$.
When plotting the velocity map with contours, see Fig. \ref{fig:v_star_log}, a twist for the line of zero velocity becomes 
apparent, it is not aligned with the photometric minor axis.

\begin{figure}
\resizebox{\hsize}{!}{\includegraphics{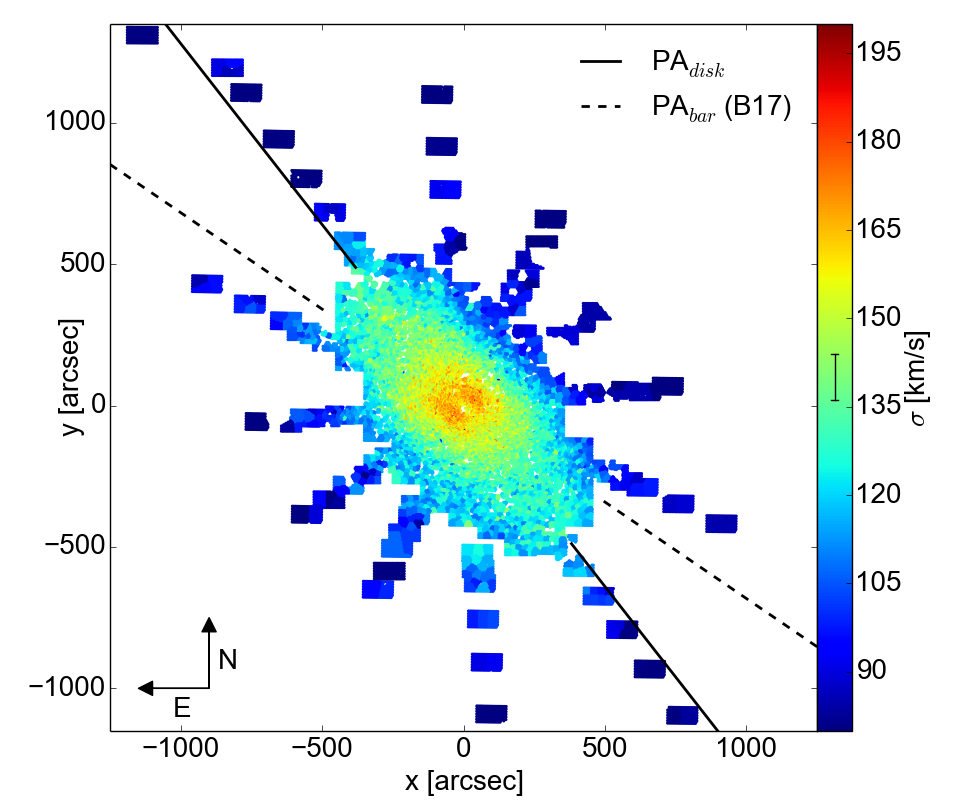}}
 \caption[Stellar velocity dispersion]{Stellar velocity dispersion map with the disk major axis with PA$_{disk}$=38$\degr$ (solid line) and the bar major axis with PA$_{bar}$=55.7$\degr$ (dashed line) from the model by \citetalias{Blana16a}.
 The line in the colorbar is the median errorbar of the velocity dispersion.}
\label{fig:sigma_star}
\end{figure}

The stellar velocity dispersion of M31 is plotted in Fig. \ref{fig:sigma_star}.
In the center, there are two lobes with high values of velocity dispersion of over 165 km s$^{-1}$ (red regions in 
Fig. \ref{fig:sigma_star}), which are separated by 
a line of lower velocity dispersion along the major axis of the bar. Around this we see a region with slightly lower 
values ($\approx$ 160 km s$^{-1}$) that has the shape of a small parallelogram (yellow area in Fig. \ref{fig:sigma_star}). Further out, we measure a region with even lower values of $\sigma$ ($ \approx$ 130 km s$^{-1}$), which is elongated along the disk major axis (light blue area in 
Fig. \ref{fig:sigma_star}). $\sigma$ drops faster along the disk minor axis than the major axis. The maximum value is $\sigma_{max}$ = 188 $\pm$ 5 km s$^{-1}$, located at a distance of 46\arcsec from the 
center, the minimum value is $\sigma_{min}$ = 55 $\pm$ 4 km s$^{-1}$, in the outermost northern disk pointing 
at x=-100\arcsec and y=1100\arcsec. The mean velocity dispersion for 
the whole dataset is $\sigma_{mean}$ = 116 $\pm$ 4 km s$^{-1}$, in the central 20\arcsec  it is $\sigma_{central}=160 \pm 5$ km s$^{-1}$, considering only the bulge region 
it is $\sigma_{mean, bulge}$=137 $\pm$ 4 km s$^{-1}$, in the disk it is considerably lower with $\sigma_{mean, disk}$=103 $\pm$ 4 km s$^{-1}$.
The disk velocity dispersion is still quite high, which is in 
agreement with what \citet{Fabricius12c} have found for other disk galaxies.

\begin{figure}
 \resizebox{\hsize}{!}{\includegraphics{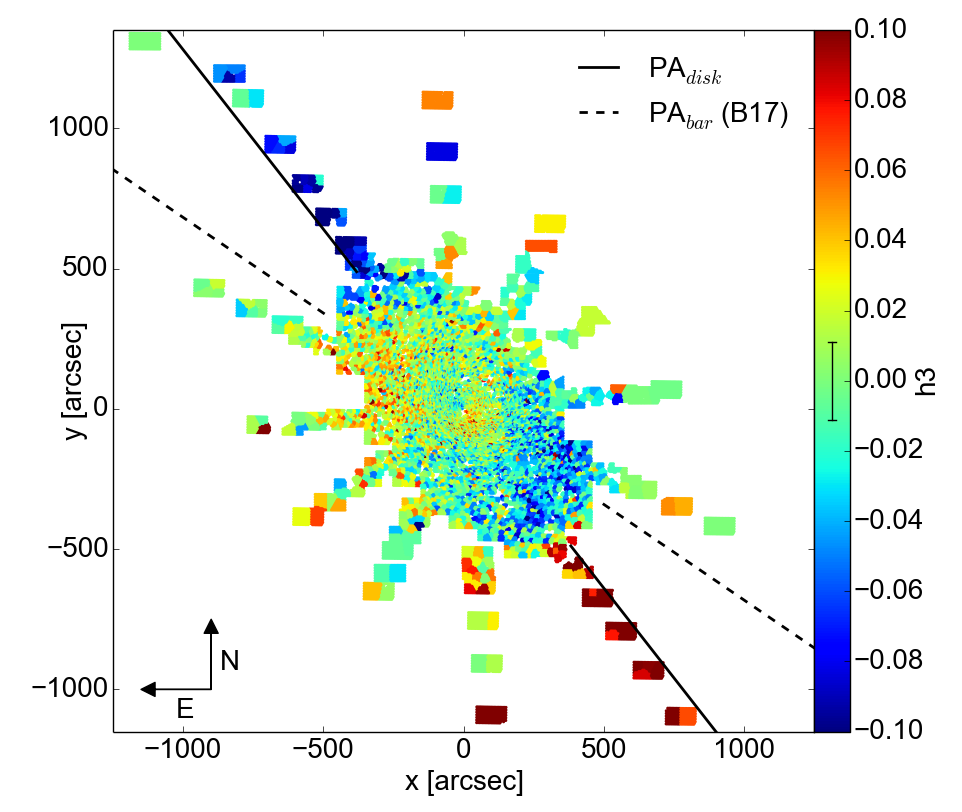}}
 \caption[h3 moment]{Map of $h3$, the third moment of the Gauss-Hermite series. The lines are analogous to Fig. \ref{fig:sigma_star}.} 
 \label{fig:h3}
 \end{figure}
In Fig. \ref{fig:h3}, we plot the Gauss-Hermite moment 
$h3$. In the very central 100\arcsec , $h3$ has high positive and negative values, while further out, 
the absolute values become lower. Further out in the bulge, the absolute values become higher again. In the disk along the major axis, the 
signs of $h3$ flip and even higher absolute values are reached. Corresponding regions of high and low values 
are relatively symmetric with respect to the center.\\
In the central 50\arcsec, $h3$ has values of about $\pm$0.05. Further out, the 
values drop to about $\pm$0.02, before rising at the edges of the bulge region 
to $\pm$0.08. In the bulge, the mean absolute value is $\overline{|h3_{bulge}|}$ = 0.03 $\pm$ 0.02.
In the disk, the values are higher, the mean absolute disk value is $\overline{|h3_{disk}|}$ = 0.04 $\pm$ 0.03.
Along the disk major axis, the values are highest, here, the mean absolute value is $\overline{|h3_{major}|}$ = 0.07 $\pm$ 0.02.
The maximum and minimum values of the whole map are also found along the disk major axis, being $h3_{max}$ = 0.14 $\pm$ 0.02 and $h3_{min}$ = -0.16 $\pm$ 0.02. 
Over most of the bulge region, we see a correlation between $h3$ and $v$, this will be discussed in section \ref{sec:Bar}.

\begin{figure}
\resizebox{\hsize}{!}{\includegraphics{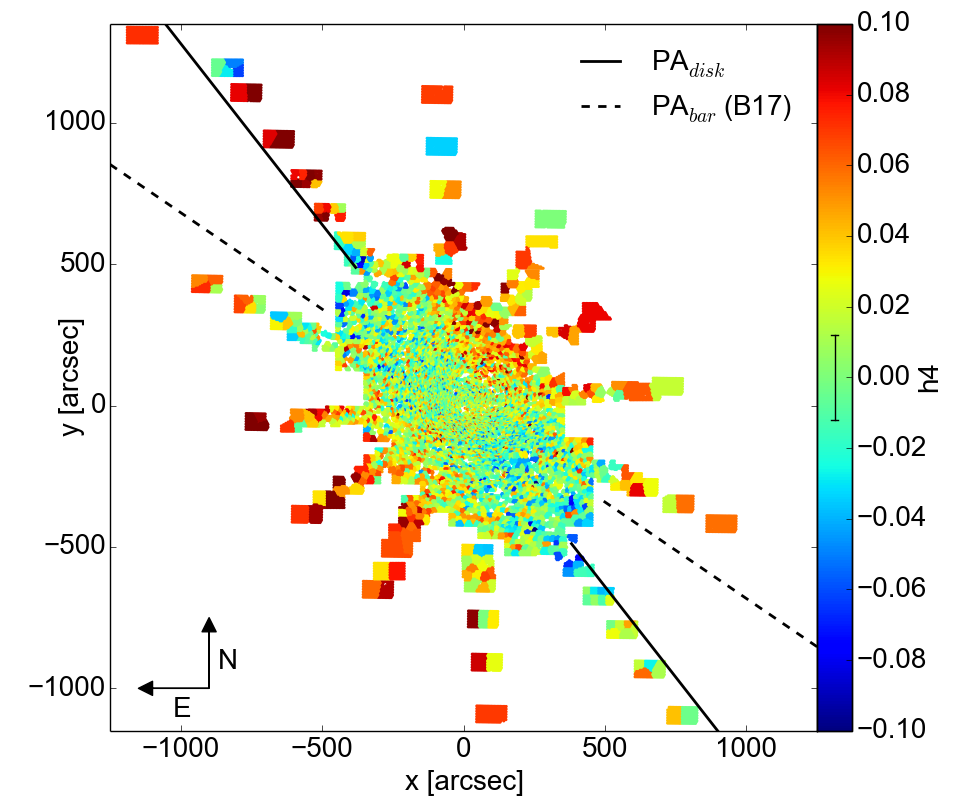}}
 \caption[h4 moment]{Map of h4, the fourth moment in the Gauss-Hermite series. The lines are analogous to Fig. \ref{fig:sigma_star}.}
 \label{fig:h4}
\end{figure}

In Fig. \ref{fig:h4}, the Gauss-Hermite moment $h4$ is presented. This map shows that along the disk major axis, the absolute values of $h4$ are very low. At the ends of the bulge region, the values of $h4$ become negative. Perpendicular to the
disk major axis, the values of $h4$ become positive. 
The mean value
over the whole dataset is $\overline{h4}=0.02 \pm 0.02$. In a 240\arcsec \ wide and 350\arcsec \ long strip along the disk major axis, the mean 
value is lower, at $\overline{h4_{outer\ bulge}}=0.002$. Along the disk minor axis, the values of $h4$ are generally 
higher, in regions that have a distance to the disk major axis of more than
120\arcsec, the mean value is $\overline{h4_{minor}}=0.03$.  In the disk 
pointings along the disk major axis, $h4$ is higher in the northern half of 
the galaxy, having a mean value of $\overline{h4_{major, north}}=0.04$, while in the southern half, it is $\overline{h4_{major, south}}=-0.01$.

\subsection{Gas kinematics}
\label{sec:GasKinematics}
\begin{figure}
\resizebox{\hsize}{!}{\includegraphics{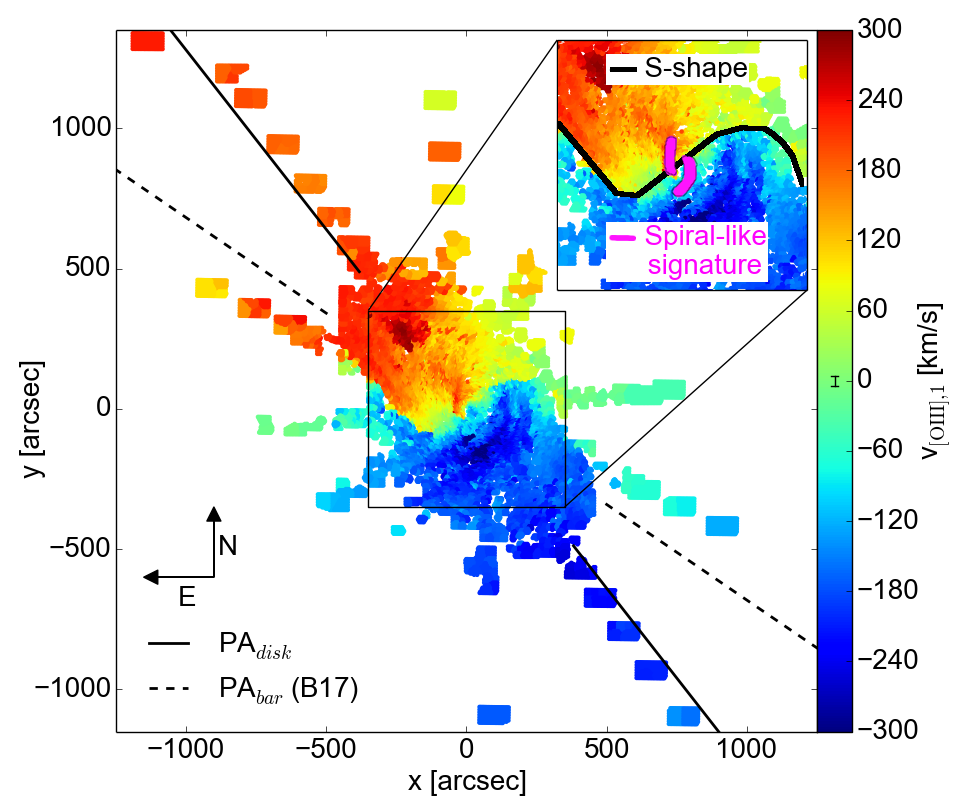}}
\caption[Velocity map of first gas component]{Velocity map of the more rapidly rotating gas component 1. The lines are analogous 
to Fig. \ref{fig:sigma_star}. The inset shows a zoom into the inner regions, with the 
``S-shape'' highlighted along the line with systemic velocity. The ``spiral-like signature'' is highlighted with the magenta 
line, which encompasses both arms of the spiral. Both these structures are mentioned in the text.}
\label{fig:first_comp}
\end{figure}

\begin{figure}
\resizebox{\hsize}{!}{\includegraphics{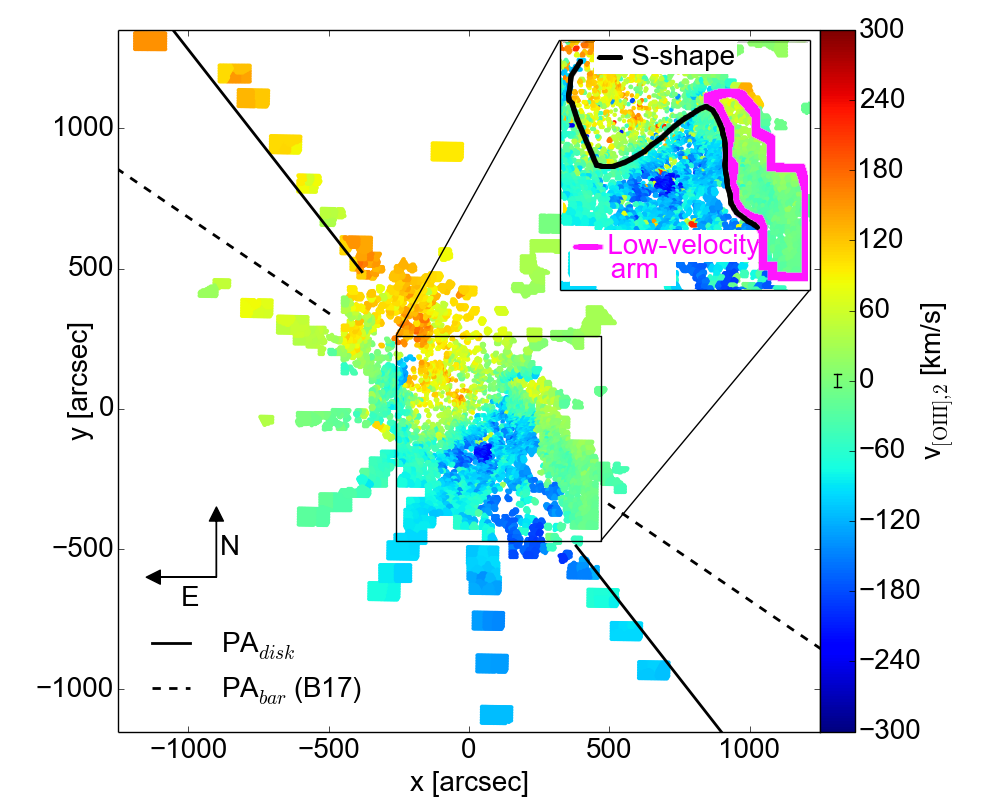}}
\caption[Velocity map of second gas component]{Velocity map of the more slowly rotating gas component 2. The lines are 
analogous to Fig. \ref{fig:sigma_star}. The inset shows a zoom into the inner regions, with the ``S-shape'' and the ``low-velocity arm''
highlighted, which are mentioned in the text.}
\label{fig:second_comp}
\end{figure}

In this section, the kinematics of the ionized gas is presented. These data will be made available online, an example table is given in table \ref{tab:Kinematic_gas_data} in the Appendix.
The motions of the gas are more complicated than the ones of the stars from 
section \ref{sec:StellarKinematics}. More complicated gas kinematics are often observed in disk galaxies (e.g. \citet{Falcon-Barroso06, Ganda06}), because contrary to the dissipationless stars, 
the gas can also lose energy through radiation. 
The dense gas traced by ground-state CO and HI transitions is most likely to have settled 
onto closed orbits via hydrodynamic interactions. Associated with this dense gas 
are regions of ionized gas \citep{Stark94}, which is then seen in the optical
emission lines. 
As mentioned in section \ref{sec:Observations}, we see in about half of the investigated binned 
spectra two gas components separated in velocity. We tried several ways to sort the two components, e.g. 
sorting the components by flux or proximity in position space. In the 
end, we settled on sorting by velocity, because it produced the smoothest maps.
The velocity map of the first component, which is the faster one of the two, is plotted in Fig. \ref{fig:first_comp}, the one of the second component in Fig. \ref{fig:second_comp}. 
The first gas component has a median absolute value of $\overline{|v_{[OIII,1]}|}$=162 $\pm$ 5 km s$^{-1}$ . 
The maximum value is $v_{[OIII,1],max}=294.7$ $\pm$ 4.5 km s$^{-1}$  at the coordinates (-220\arcsec, 281\arcsec), at 
360\arcsec from the center along the disk major axis in the receding side. The minimum value is $v_{[OIII,1],min}=-340$ $\pm$ 3.0 km s$^{-1}$  at the coordinates (25\arcsec,-100\arcsec), which is 
at about 100\arcsec south of the center in the approaching side, is closer to the center
than the maximum on the other side, leading to an asymmetric appearance with respect to the center. Apart from that, there is a ``spiral-like signature'' 
in the innermost 100\arcsec $\times$ 100\arcsec and a large ``S-shape'' between the approaching and receding 
gas velocities. These structures are marked in the zoomed-in inset of figure \ref{fig:first_comp}. 

For the second component, the median absolute value is significantly lower than for the first component, 
with $\overline{|v_{[OIII,2]}|}$=73.2 $\pm$ 5.5 km s$^{-1}$.
The maximum value is $v_{[OIII,2],max}=183.8$ $\pm$ 5.2 km s$^{-1}$  at the coordinates (-190\arcsec, 280\arcsec), which is in the same region 
as the maximum for the first component. The minimum value is $v_{[OIII,2],min}=-240.2$ $\pm$ 3.8 km s$^{-1}$, at the coordinates 
(65\arcsec, -170\arcsec), at 182\arcsec from the center along the disk major axis in the receding side. 
This again corresponds to a region of low velocities in the first component. The overall shape 
of the velocity field for the second component is similar to the first, also with an ``S-shape'' along the line of 
zero velocity. This structure has a slightly different shape than the one 
for the first component, an arm of approaching velocities is 
extending further into the region of receding velocities. Additionally, on the western side of the bulge, at about 500\arcsec along the disk major axis from the center, there is an arm of velocities of about zero. This arm seems to be 
unconnected to the kinematics of the rest of this component. These structures are marked in the zoomed-in inset of Fig. \ref{fig:second_comp}

\begin{figure}
\begin{minipage}{0.49\textwidth}
\includegraphics[width=\textwidth]{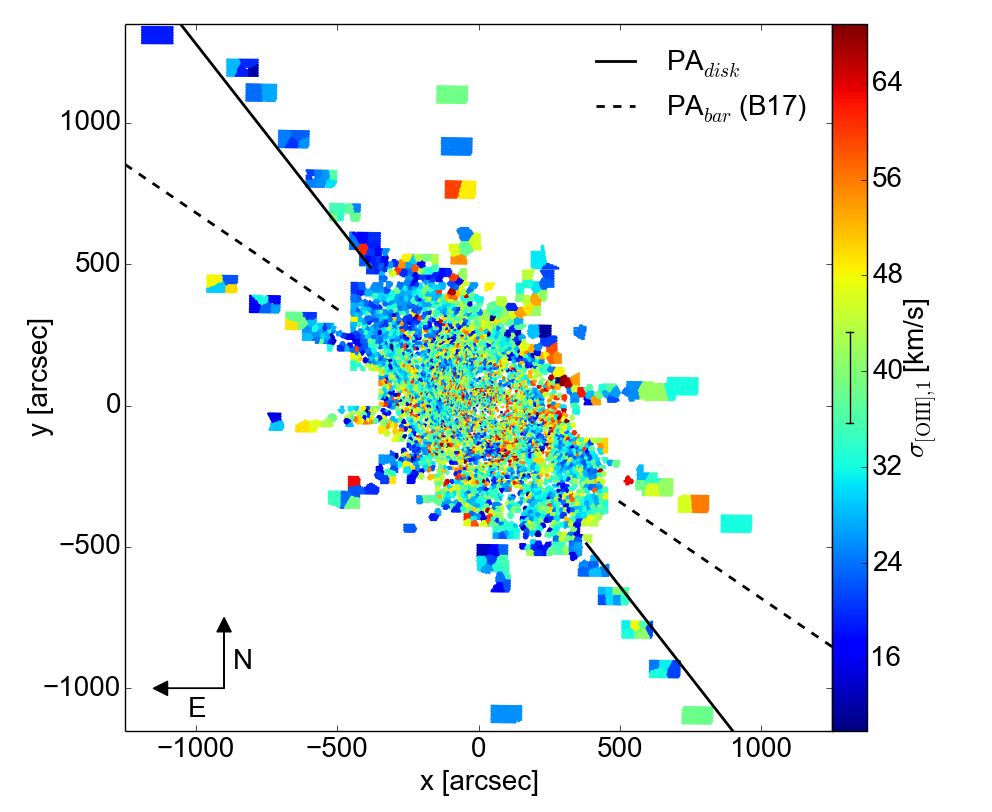}
\caption[Velocity dispersion of first gas component]{Velocity dispersion of the first gas component. The lines are analogous to Fig. \ref{fig:sigma_star}.}
\label{fig:sigma_gas_1}
\end{minipage}
\hfill
\begin{minipage}{0.49\textwidth}
\includegraphics[width=\textwidth]{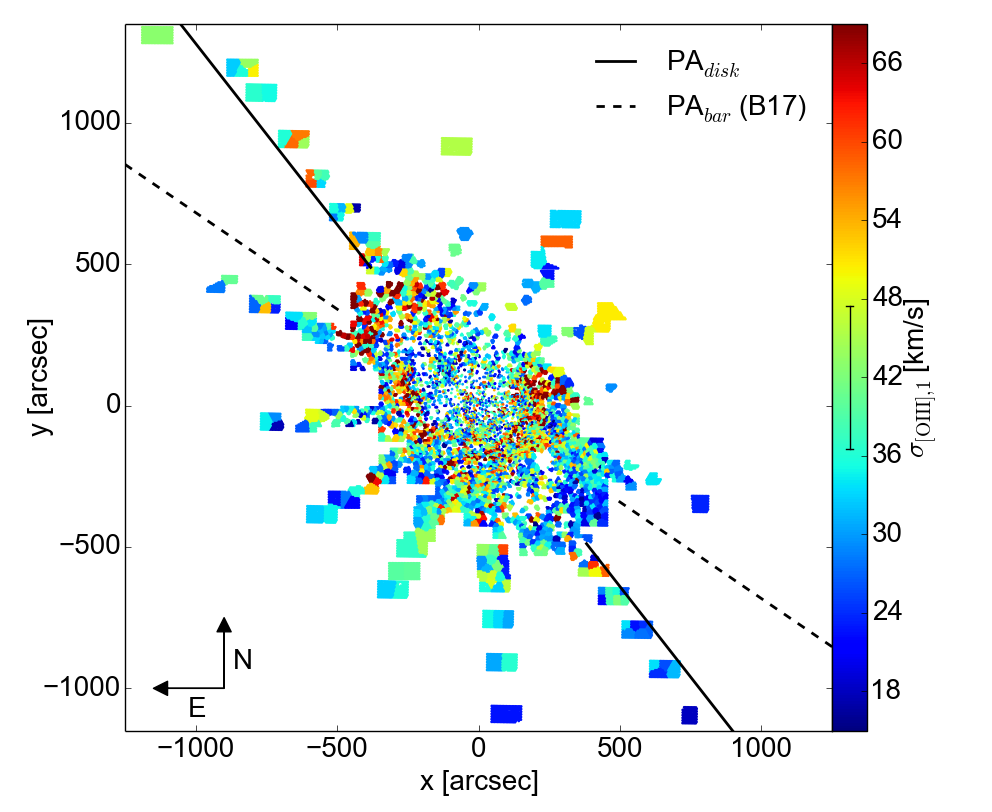}
\caption[Velocity dispersion of second gas component]{Velocity dispersion of the second gas component. The lines are analogous to Fig. \ref{fig:sigma_star}.}
\label{fig:sigma_gas_2}
\end{minipage}
\end{figure}
 The velocity dispersion of the first component is plotted in Fig. \ref{fig:sigma_gas_1}. The general appearance is noisy, overall, the 
 mean value is $\overline{\sigma_{OIII,1}} = 35 \pm 5$ km s$^{-1}$. The velocity dispersion is low near the northern end of the bulge along the major axis, where the velocities are high ($\sigma$ $\approx$ 50 - 60 km s$^{-1}$). \\
 In the velocity dispersion values of the second component, a ring-like structure of high values is visible, which increases the 
 mean value of the velocity dispersion to $\overline{\sigma_{OIII,2}}=37 \pm 6$ km s$^{-1}$. The values in the ring are  $\overline{\sigma_{OIII,2}} = 57 \pm 10$ km s$^{-1}$. 
 This ring of high velocity dispersions corresponds to a similar ring in Fig. \ref{fig:nlines}, which is where two peaks are measured.  
 In this ring, a second peak with a lower amplitude than the first peak is present. However, this second peak is broader, therefore the velocity dispersion is 
 larger.  

\subsection{Gas fluxes}
\label{sec:flux}
\begin{figure}
\resizebox{\hsize}{!}{\includegraphics{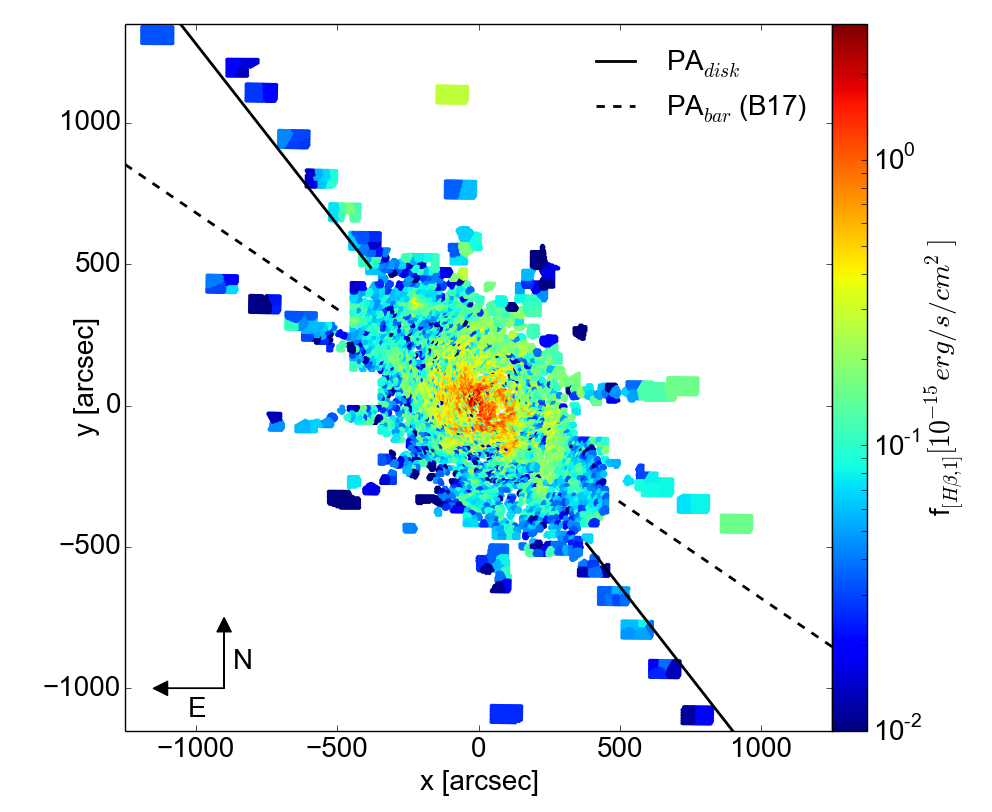}}
 \caption[Flux of first H$\beta$ component]{Flux of first H$\beta$ component. The median error is 0.01$\cdot 10^{-15}$ [erg/s/cm$^2$]. The lines are analogous to Fig. \ref{fig:sigma_star}.}
 \label{fig:flux_Hb_1}
 \end{figure}

\begin{figure}
 \resizebox{\hsize}{!}{\includegraphics{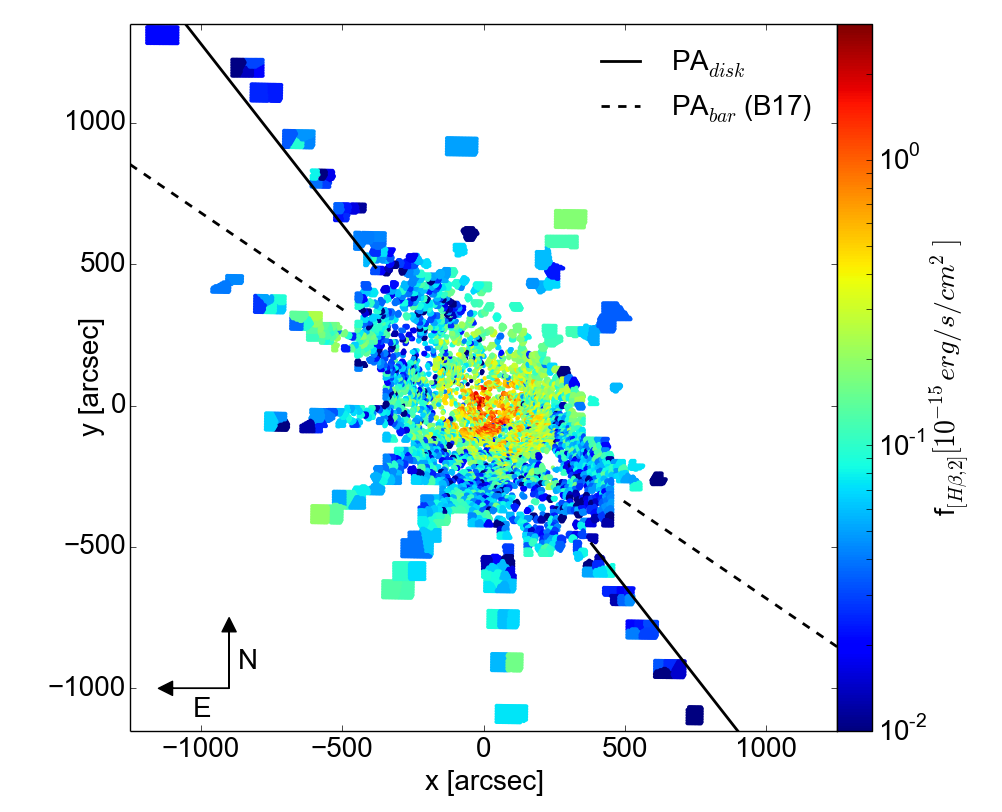}}
\caption[Flux of second H$\beta$ component]{Flux of second H$\beta$ component. The median error is 0.01$\cdot 10^{-15}$ [erg/s/cm$^2$]. The lines are analogous to Fig. \ref{fig:sigma_star}.}
\label{fig:flux_Hb_2}
\end{figure}

\begin{figure}
\resizebox{\hsize}{!}{\includegraphics{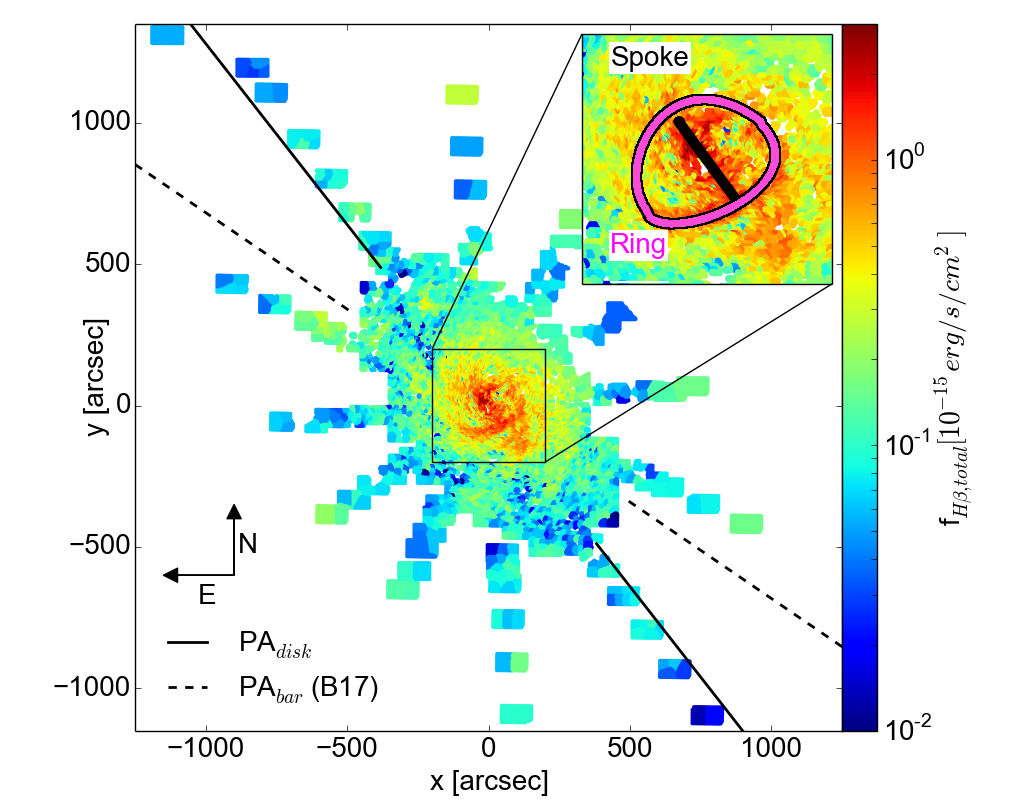}}
\caption[Flux of both H$\beta$ components combined]{Flux of both H$\beta$ components combined. The median error is 0.02$\cdot 10^{-15}$ [erg/s/cm$^2$]. The lines are analogous to Fig. \ref{fig:sigma_star}. The inset shows a zoom into the inner 200\arcsec$\times$200\arcsec, 
with the structures ``ring'' and ``spoke'' highlighted, which are mentioned in the text.}
\label{fig:flux_Hb_combined}
\end{figure}

\begin{figure}
\resizebox{\hsize}{!}{\includegraphics{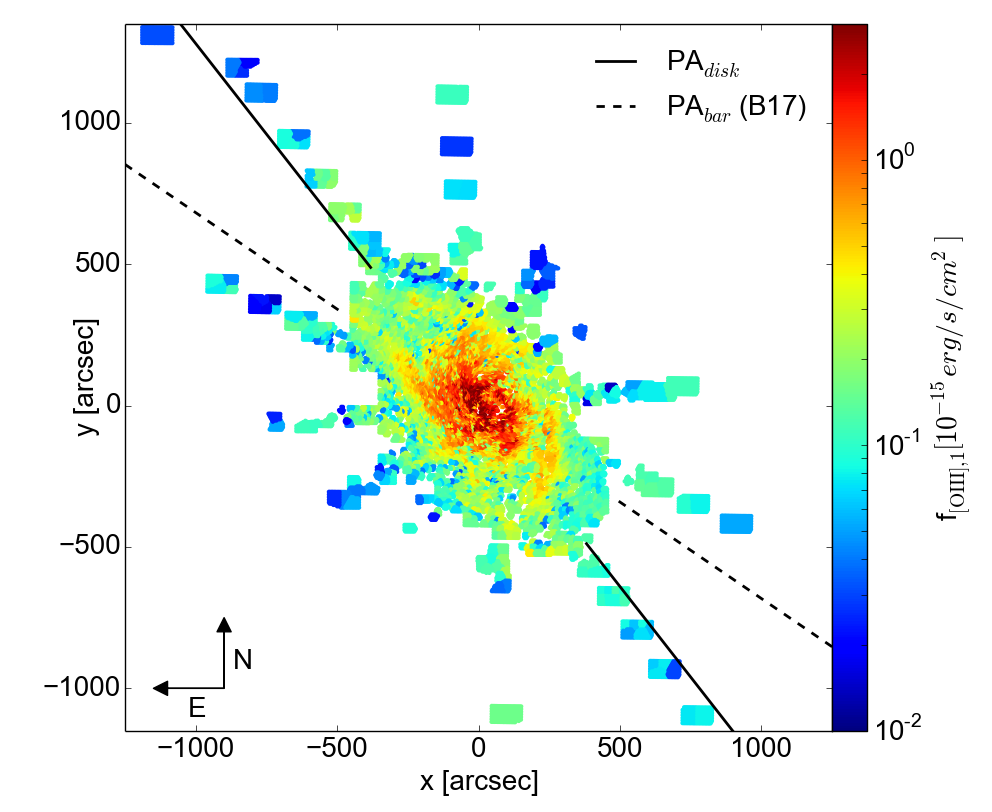}}
 \caption[Flux of first {[OIII]} component]{Flux of first [OIII] component. The median error is 0.01$\cdot 10^{-15}$ [erg/s/cm$^2$]. The lines are analogous to Fig. \ref{fig:sigma_star}.} 
 \label{fig:flux_OIII_1}
  \end{figure}

\begin{figure}
\resizebox{\hsize}{!}{\includegraphics{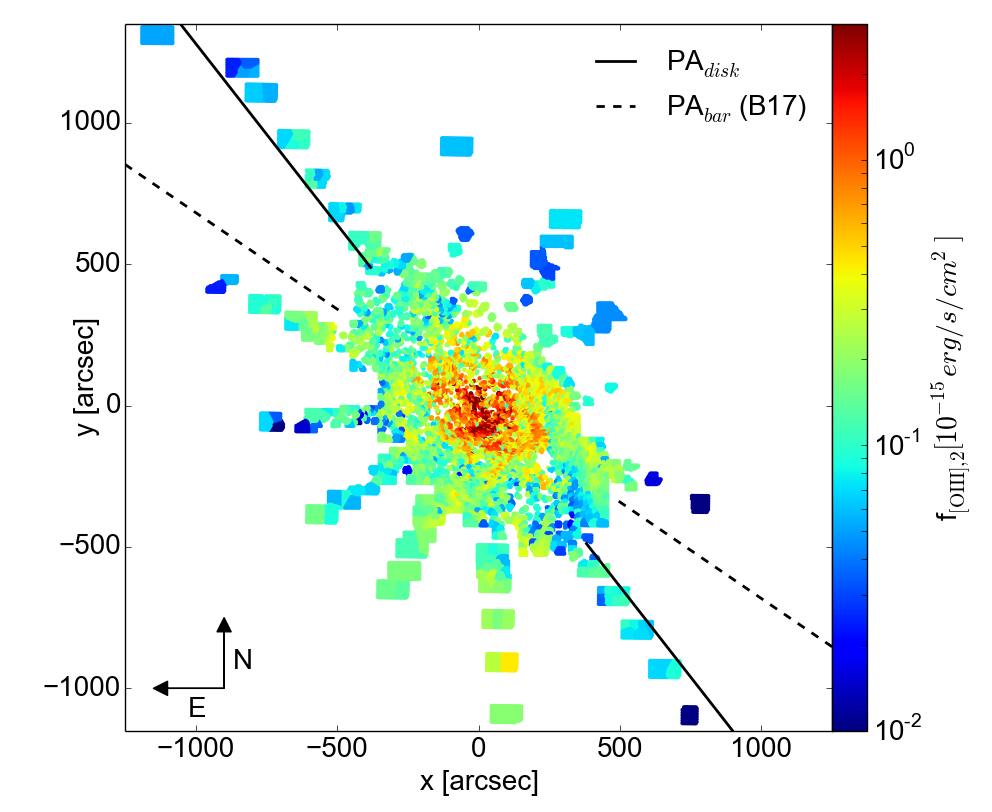}}
\caption[Flux of the second {[OIII]} component]{Flux of the second [OIII] component. The median error is 0.01$\cdot 10^{-15}$ [erg/s/cm$^2$]. The lines are analogous to Fig. \ref{fig:sigma_star}.}
\label{fig:flux_OIII_2}
\end{figure}

\begin{figure}
\resizebox{\hsize}{!}{\includegraphics{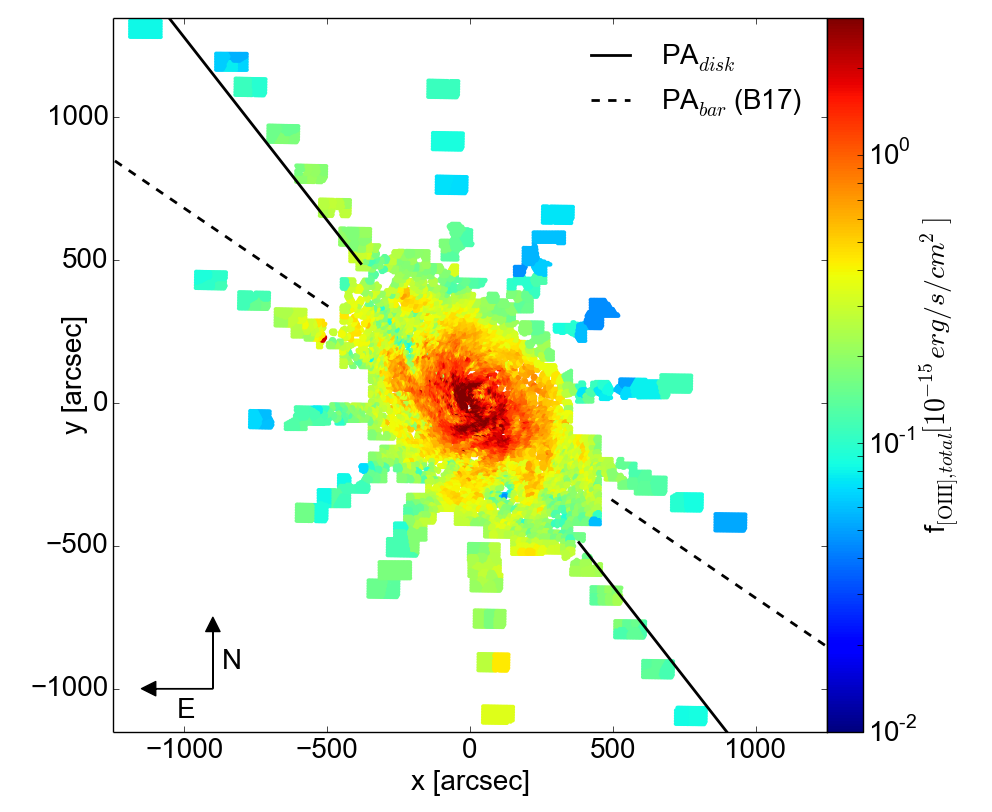}}
\caption[Flux of both {[OIII]} components combined]{Sum of the flux of the two [OIII] components. The median error is 0.02$\cdot 10^{-15}$ [erg/s/cm$^2$]. The lines are analogous to Fig. \ref{fig:sigma_star}.}
\label{fig:flux_OIII_combined}
\end{figure}

\begin{figure}
\resizebox{\hsize}{!}{\includegraphics{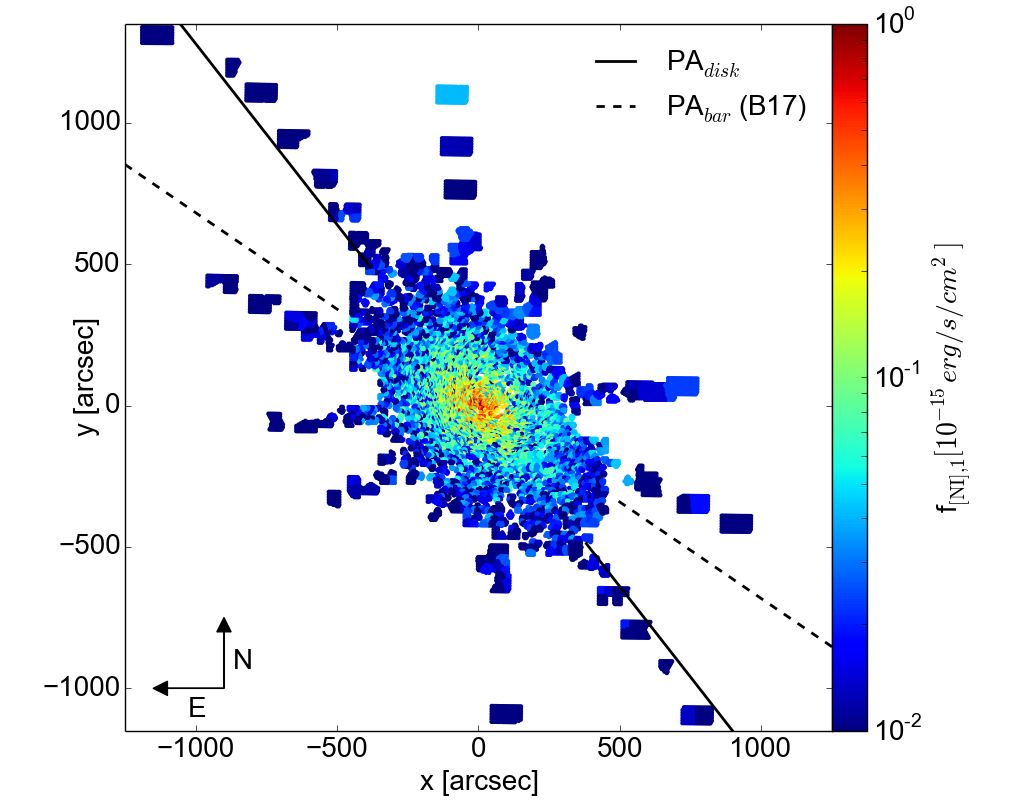}}
 \caption[Flux of the first {[NI]} component]{Flux of the first [NI] component. The median error is 0.01$\cdot 10^{-15}$ [erg/s/cm$^2$]. The lines are analogous to Fig. \ref{fig:sigma_star}.}
\label{fig:flux_NI_1}
 \end{figure}

\begin{figure}
 \resizebox{\hsize}{!}{\includegraphics{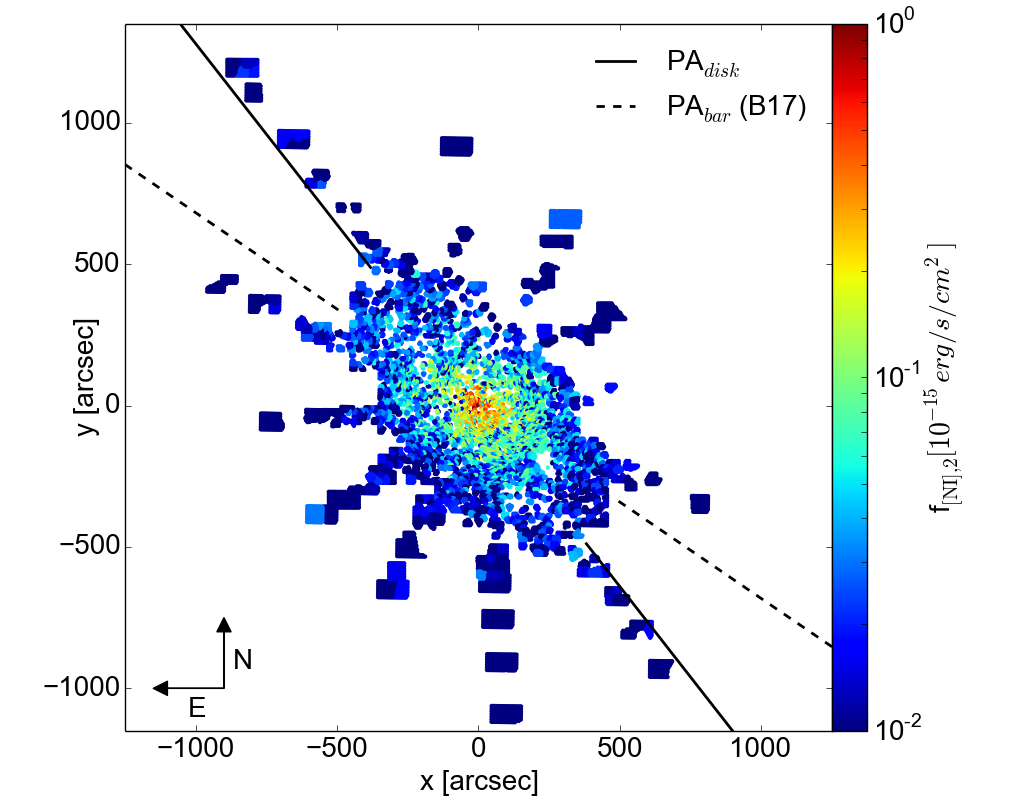}}
\caption[Flux of the second {[NI]} component]{Flux of the second [NI] component. The median error is 0.01$\cdot 10^{-15}$ [erg/s/cm$^2$]. The lines are analogous to Fig. \ref{fig:sigma_star}.}
\label{fig:flux_NI_2}
\end{figure}

\begin{figure}
\resizebox{\hsize}{!}{\includegraphics{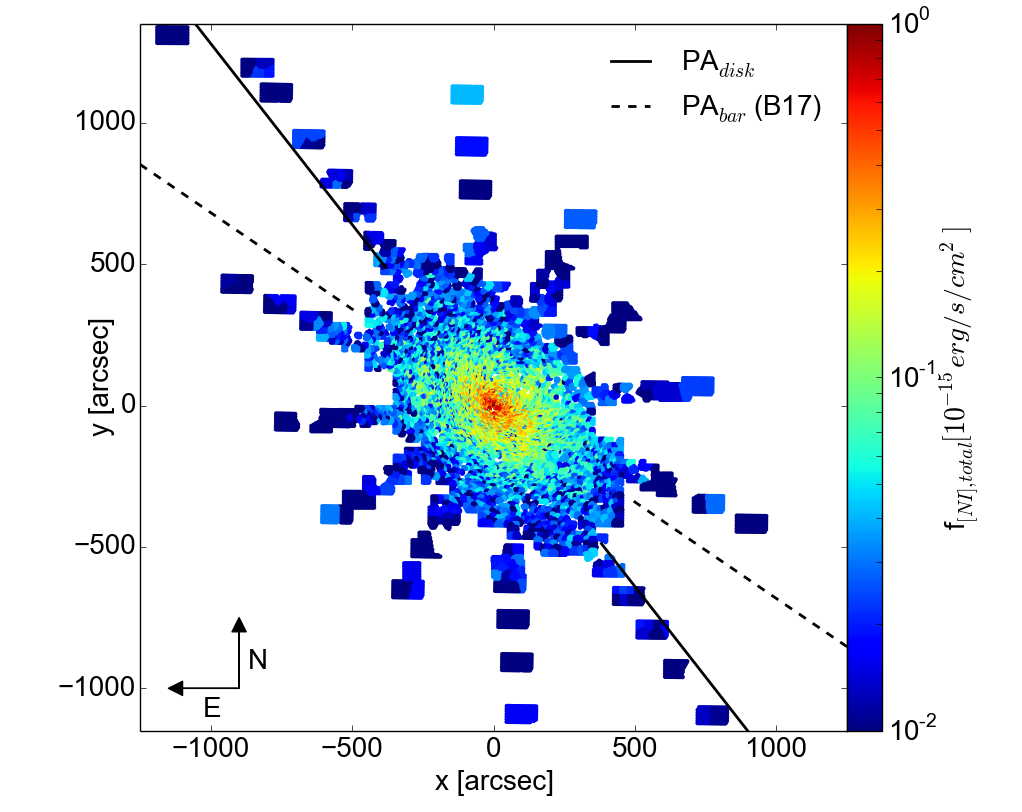}}
\caption[Flux of both {[NI]} components combined]{Total flux of the two [NI] components. The median errorbar is 0.01$\cdot 10^{-15}$ [erg/s/cm$^2$]. The lines are analogous to Fig. \ref{fig:sigma_star}.}
\label{fig:flux_NI_combined}
\end{figure}

The fluxes of H$\beta$, [OIII]$\lambda$5007 and [NI]$\lambda$5198 are plotted in Figs. \ref{fig:flux_Hb_1} to \ref{fig:flux_NI_combined}. 
The values will be made available online, an example table is given in Table \ref{tab:Photometric_gas_data} in the Appendix.
In Fig. \ref{fig:flux_Hb_1} and Fig. \ref{fig:flux_Hb_2}, we show the line flux of the first and second H$\beta$ line, while in Fig. \ref{fig:flux_Hb_combined}, we show the 
sum of the two components. The corresponding fluxes for [OIII]$\lambda$5007 are shown in Fig. \ref{fig:flux_OIII_1}, Fig. \ref{fig:flux_OIII_2}, and Fig. \ref{fig:flux_OIII_combined}.  The plots for [NI]$\lambda$5198
are plotted in Fig. \ref{fig:flux_NI_1}, Fig. \ref{fig:flux_NI_2} and Fig. \ref{fig:flux_NI_combined}. \\
There are regions where the first component has higher flux than the second one, as well as regions where the opposite is true. When averaging the fluxes along ellipses, however, it becomes apparent that 
the first component is stronger than the second one. This is true for all different lines.
In the H$\beta$ and [OIII] maps, a spiral structure is visible over most of the bulge region, with an incomplete elliptical ring 
inside, which is oriented roughly along the minor axis, with a ``spoke'' along the ring's short axis. The ring and spoke are 
highlighted in the zoomed-in inset of Fig. \ref{fig:flux_Hb_combined}. 
This ring is at a smaller radius than the one mentioned above in the velocity dispersion of the second gas component.
In the southwest, one of the arms of the spiral pattern is prominent in the H$\beta$ and the [OIII]. 
The fluxes of H$\beta$ are lower than the ones of [OIII]. The [NI] is much fainter than either the H$\beta$ or the [OIII], no clear pattern can be seen there apart from the 
fact that it becomes brighter in the center. The overall filamentary appearance of the gas morphology could be either due to heating by shocks or 
supernovae of type Ia \citep{Jacoby85}.


\begin{figure}
 \resizebox{\hsize}{!}{\includegraphics{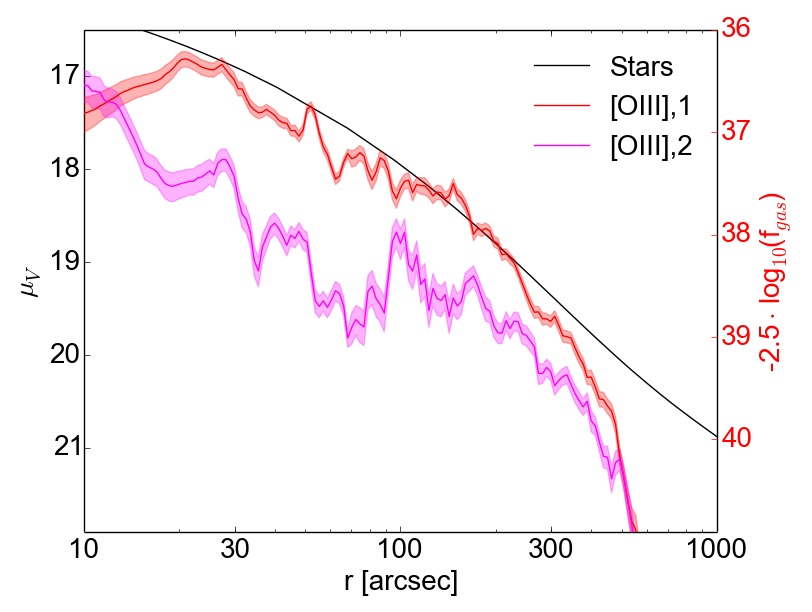}}
 \caption[Comparison stellar and gas surface brightness]{Comparison of the surface brightness of the stars (black) taken from \citet{Kormendy99} 
 with ellipse fits to the fluxes of the first [OIII] component (red) and the second [OIII] component (magenta). The left y-axis is valid for the stellar surface brightness, 
 the one on the right for the fluxes of the two gas components. The scales have been adjusted so that the profiles for the 
 stars and the first [OIII] component can be compared easily.}
 \label{fig:comparison_stars_gas}
\end{figure}

We did ellipse fits to the maps of the [OIII] gas and compared it to the stellar surface brightness. 
The comparison is shown in Fig. \ref{fig:comparison_stars_gas}. 
From about 30\arcsec to about 200\arcsec, the decline of the [OIII] is similar to the one of the stars, while further out, 
it drops significantly compared to the light of the stars. This corresponds to the edge of the 
spiral structure visible in the maps. The second [OIII] component has lower flux values, its profile is 
more irregular, but it also declines fast from 200\arcsec \ outwards. 

 \section{Bar signatures in kinematics and morphology}
 \label{sec:Bar}
 \subsection{Bar signatures in the stellar kinematics}
A bar leaves certain signatures in the kinematics of the stars. \citet{Bureau05} modeled several bars, with different strengths and 
viewing angles. \citet{Stark94}, \citet{Athanassoula06} and \citetalias{Blana16a} all find that the bar in M31
is neither viewed end-on nor side-on, but instead at an intermediate angle. 
The special signatures of a bar are plateaus in the velocity and the velocity dispersion, and minima in the higher moments $h3$ and $h4$. These were theoretically predicted 
by \citet{Bureau05} and measured on several barred galaxies by \citet{Chung04}.
We adopt the model of \citetalias{Blana16a} as the one against which we compare our results. In their model, the bar has 
a position angle of PA$_{bar}$=55.7$\degr$, extending out to 600\arcsec \ in projection.
However, in the direction of the bar, our coverage only extends to 500\arcsec, we are therefore missing 
the crucial regions of the end of the bar. Pointings covering these regions are being observed
and the results will be presented in a future paper. The co-rotation radius of \citetalias{Blana16a} and the outer 
Lindblad resonance are also outside our observed regions.

\begin{figure}
 \resizebox{\hsize}{!}{\includegraphics{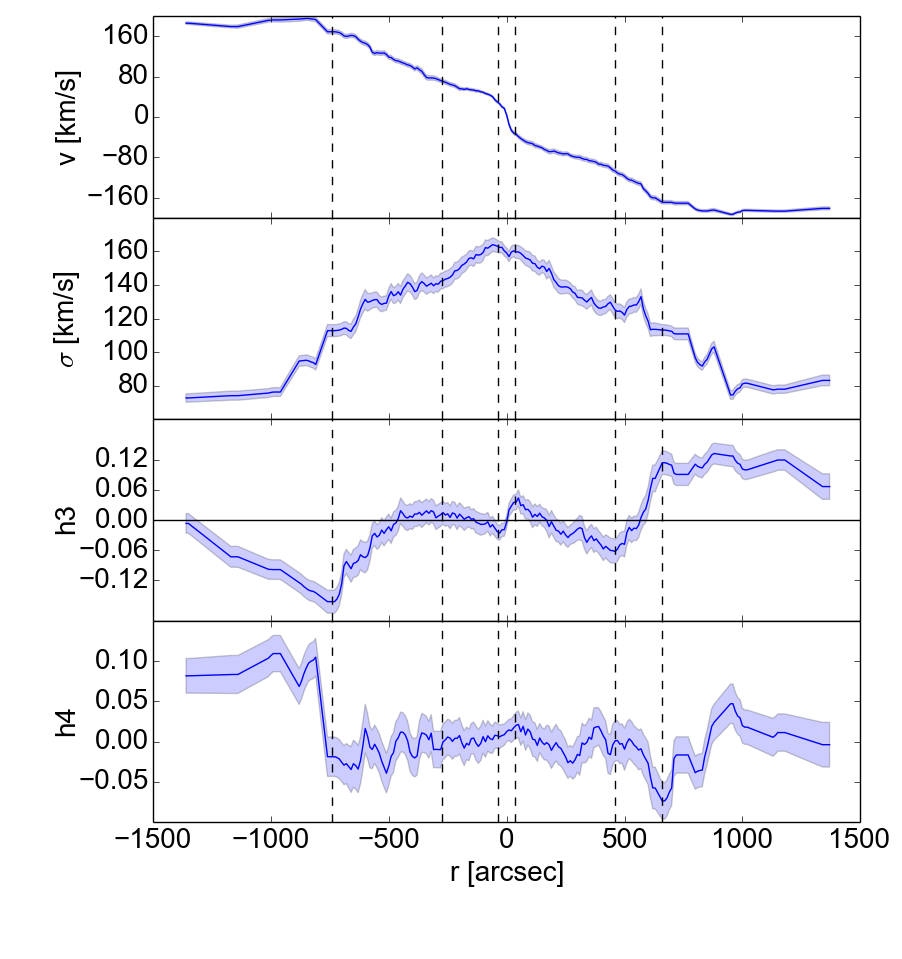}}
 \caption[Cuts along the disk major axis for the stellar velocity]{Cuts through the 
 stellar kinematic maps from \texttt{VIRUS-W} along the disk major axis (PA=38$\degr$). In the third panel, the horizontal line is $h3$=0. The 
 vertical dashed lines highlight the position where $h3$ has a local maximum or minimum.}
 \label{fig:own_cuts}
\end{figure}
In Fig. \ref{fig:own_cuts}, cuts through the kinematic fields along the disk major axis are plotted. Left corresponds to the north-eastern receding side, right to the south-western 
approaching side.
The velocity rises to 70 km s$^{-1}$  at 35\arcsec, before remaining relatively constant 
until 350\arcsec on the left and 460\arcsec on the right, before reaching 160 km s$^{-1}$  at -750\arcsec on the left and at 660\arcsec on the right. 
An axisymmetric stellar disk rises rapidly, but smoothly and remains flat at large radii.
\citet{Bureau05} found that a bar seen end-on produces a ''double-hump`` profile, where the velocity rises rapidly to a local 
maximum, then drops slightly to create a local minimum and then rises again slowly to the flat section at the end. 
When the bar is not seen end-on, but at an intermediate angle, this behavior is weakened and the local maximum and minimum disappear 
and form a single plateau at moderate radii, which is what we see in the profiles of M31. This is expected from the models by \citetalias{Blana16a}, where 
the bar is seen at an intermediate angle.
The rising part of the velocity curve comes from the orbits of the stars that are parallel to the bar, while the slower growth afterwards, 
or the local minimum in the case of an end-on bar, is caused by an inner ring structure caused by the bar \citep{Bureau05}.

The velocity dispersion profile in Fig. \ref{fig:own_cuts} shows that
$\sigma$ has two off-centered maxima, with a slight drop of about 8 km s$^{-1}$ in between. {This drop was first observed by \citet{Kormendy88}, who saw them as evidence for a central disk
and subsequently the supermassive black hole.}
\citet{Bureau05} find that such minima can also be caused by the bar, because the orbits parallel to the bar become 
more circular and thus lower the dispersion, however, the minimum in models by \citet{Bureau05} is usually much stronger than what we measure here.
{One should keep in mind that such minima are not uniquely related to bars. \citet{Comeron08} suggested that cold gas is accreted to the center and 
a subsequent starburst creates then stars with low velocity dispersion. They claim that the gas is most probably driven into the center 
by spiral arms, but also, albeit less probable, bars.} \\
An inner disk can also lead to this minimum, such a structure has been 
found in the very center of M31 \citep{Tremaine95, Peiris03}, however, with a scalelength of only about 1\arcsec, this structure 
is not resolved by our observations. The drop could also be caused by the classical bulge \citepalias{Blana16a}. 
$\sigma$ drops to 140 km s$^{-1}$  at 400 \arcsec, before staying roughly constant out to 600\arcsec and then
reaching 80 km s$^{-1}$  at 950\arcsec.  These plateaus, which are not completely flat, but only less inclined than the rest of the profile, 
are also frequently seen in barred galaxies \citep{Bureau05}.

The $h3$ profile in Fig. \ref{fig:own_cuts} shows that the slope of $h3$ changes sign several times.
The maxima and minima in the $h3$ correspond to the points where the slope of the 
velocity profile changes, this is similar to the behavior of the simulations by \citet{Bureau05} and has also 
been observed 
in barred galaxies \citep{Fisher97, Chung04, Falcon-Barroso06, Ganda06}.
 \begin{figure}
 \resizebox{\hsize}{!}{\includegraphics{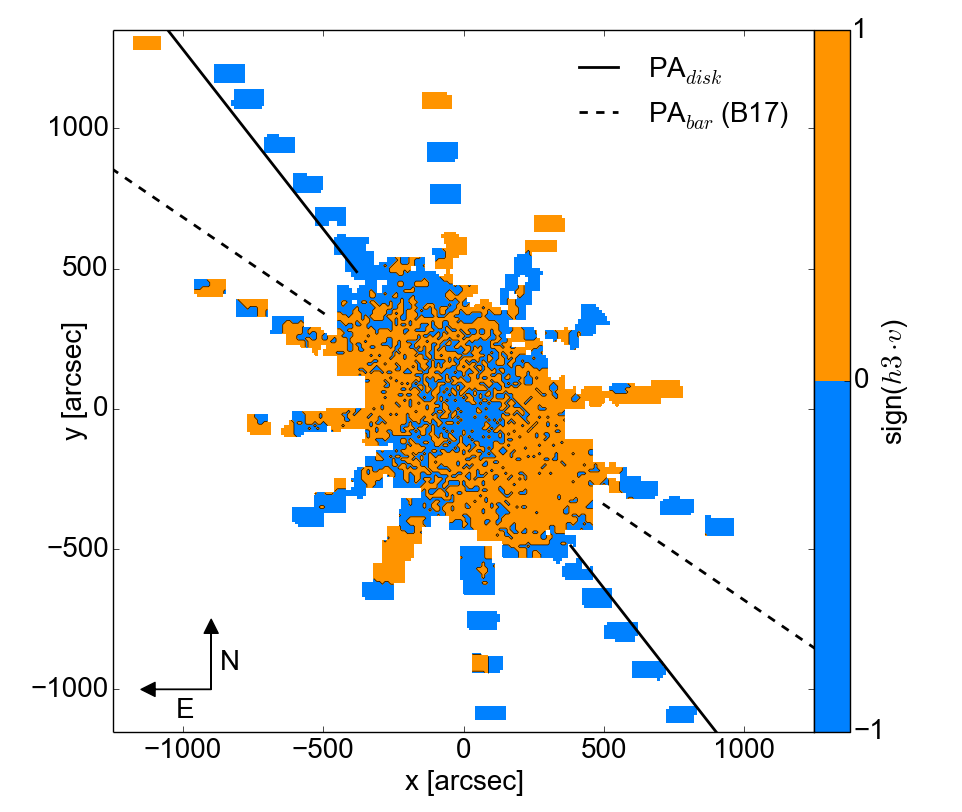}}
 \caption[Correlation between v and h3]{Correlation between the stellar velocity and $h3$. Plotted is $sign(h3 \cdot v)$, with the disk and bar major axis from Fig. \ref{fig:sigma_star}.}
 \label{fig:h3correlation}
\end{figure}
In Fig. \ref{fig:h3correlation}, we show the correlation between $h3$ and $v$. The color blue 
represents where the two quantities are anti-correlated, while in orange areas, they are correlated. 
In the outer regions, $h3$ is anti-correlated to $v$, which is the 
expected behavior for disky structures in a galaxy \citep{Binney87,Bender94, Fisher97, Binney08}.
In the central 30\arcsec, $h3$ is again anti-correlated to the velocity. This could mean that a disky structure is also present at the center, 
potentially explaining the slight drop in velocity dispersion. The radial extent of this central anti-correlation corresponds roughly to the 
rapidly rising part of the rotation curve, a behavior that is also seen in 
other disk galaxies \citep{Chung04} and interpreted as a decoupled inner disk. In between the two anti-correlated regions, $h3$ becomes correlated 
with the velocity, which is taken by \citet{Bureau05} to be a sign for a bar.
The correlated region is oriented more along the bar major axis than the disk major axis. It is more prominent in the south-west than in the north, because the northern edge of the 
bulge is more affected by dust \citep{Draine14}.
The correlation means that there are more stars rotating faster than 
the circular velocity in projection, which may be a consequence of elongated motions. However, the correlation does not necessarily have to be 
caused by a bar, it can also be caused by the superposition of an axisymmetric bulge component embedded in a rotating disk, depending on the bulge-to-disk ratio. If the 
bulge is brighter than the disk, the main velocity that is seen is mainly the bulge, with the disk creating a tail 
of high-velocity material \citep{vanderMarel93, Bureau05}. 

The $h4$ profile in Fig. \ref{fig:own_cuts} is relatively constant, with the 
exception of a minimum at 670\arcsec, where $h4$ drops from 0 to -0.07. This minimum 
corresponds to the radius where $h3$ reaches its maximum. On the opposite side, the drop in $h4$ at -750\arcsec is not as pronounced.
Outside of 1000\arcsec, the values of $h4$ stay roughly constant at larger values, $\approx$ 0.01 for positive radii and $\approx$ 0.07 for negative radii.
B/P bulges often show dips in the very center in $h4$ \citep{Debattista05, MendezAbreu08}, however, this only applies to low inclinations of $i<$30$\degr$.
It is therefore not surprising that we do not see a central drop in $h4$ in M31.  

\subsection{Bar signatures in the gas kinematics and morphology}
The simulations of \citet{Athanassoula06} and \citetalias{Blana16a} do not take into account the gas content 
in M31. However, the gas content in M31 is estimated to be only $\approx$ 7$\%$ of the stellar mass \citep{Corbelli10}, so we can 
safely assume that the gas will follow the potential set by the stellar and dark matter components. A detailed discussion of the gas behavior in this potential is postponed again to a future paper \citep{Blana16b}.
As mentioned above, the velocity maps for both gas components display an S-shape in the line of zero velocity.
This S-shape is stronger than the twist
 in the stellar velocity field. Such S-shaped twists in the gas velocity are characteristic of velocity fields 
 with oval distortions or bars \citep{Bosma81}. Many barred galaxies show them, like NGC 1068 \citep{Emsellem06}, NGC 1300 \citep{Peterson80}, NGC 2683 \citep{KuziodeNaray09}, NGC 3386 \citep{Garcia-Barreto01} and NGC 5448 \citep{Fathi05}.\\
The region of very low velocities at the edge of the bulge on the near side of the galaxy
could be produced by a large scale warp in the gas. Such a warp can project small velocities from further out into 
the line-of-sight \citep{Melchior11}. \\
\begin{figure}
\resizebox{\hsize}{!}{\includegraphics{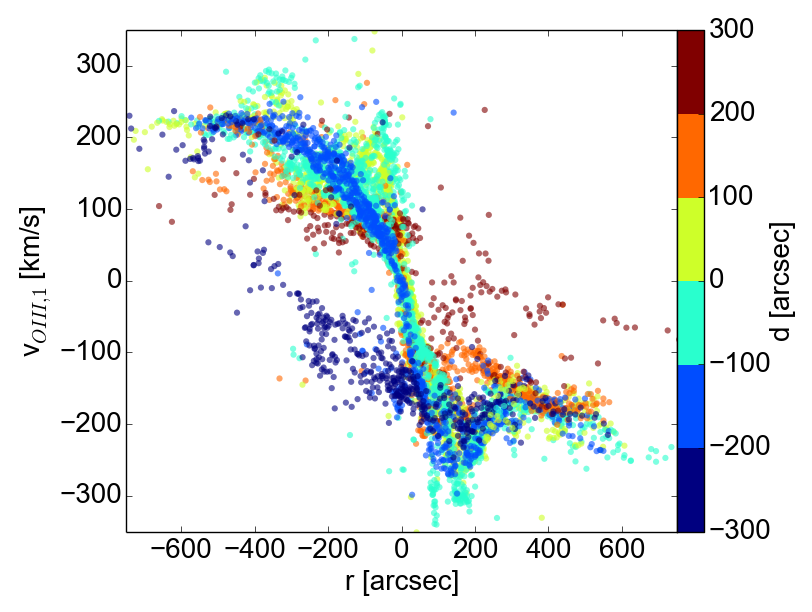}}
\caption[Position-velocity diagram of first component]{The position-velocity diagram of the first component projected onto the disk major axis, the x-axis is the distance along the major axis, the y-axis is the velocity and the 
color represents the perpendicular distance to the major axis. }
\label{fig:v_OIII_1_3d}
\end{figure}
\begin{figure}
\resizebox{\hsize}{!}{\includegraphics{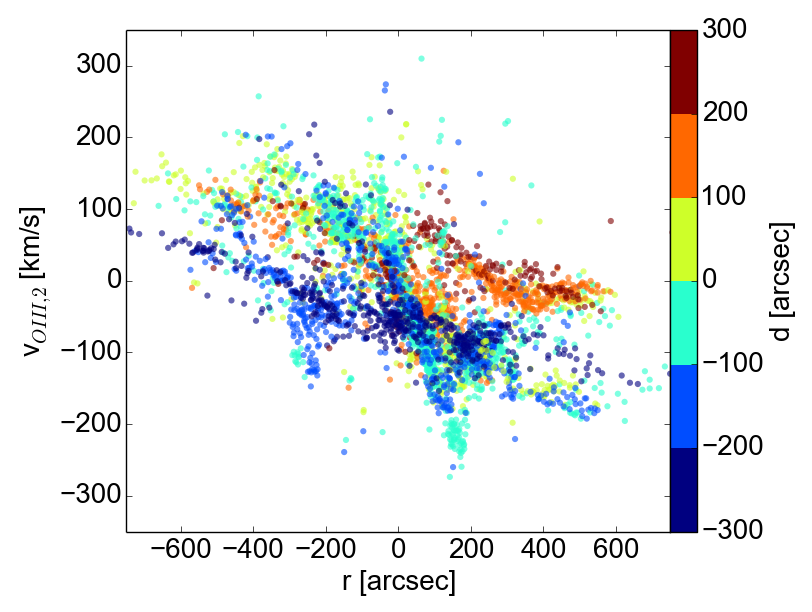}}
\caption[Position-velocity diagram of second component]{Similar to Fig. \ref{fig:v_OIII_1_3d}, the second component is plotted.}
\label{fig:v_OIII_2_3d}
\end{figure}
In Fig. \ref{fig:v_OIII_1_3d} and \ref{fig:v_OIII_2_3d}, 
position-velocity diagrams are plotted for the gas. We show the gas kinematics in this way to compare with similar diagrams from measurements and simulations of neutral gas, e.g. \citet{Chemin09} and \citet{Athanassoula06}.
We cut our maps along the disk major axis at PA=38$\degr$. At each radius, all points perpendicular to the major axis are plotted. 
They are color coded, depending on their distance to the major axis along the perpendicular coordinate d. Values on the far side of the major axis are 
shown in blue, values on the near side in red.
Fig. \ref{fig:v_OIII_1_3d} shows the first component and Fig. \ref{fig:v_OIII_2_3d} the second one. 
In this way, differences between the two components become immediately apparent. Both components occupy different regions in the PV-diagrams. The two components are real as they 
have different velocities. We also compared these PV-diagrams to the ones of the planetary nebulae from \citet{Merrett06}. As expected, neither gas component corresponds to the PNe, since the PNe follow the kinematics 
of stars.

In the PV-diagram of the first gas component,
there is a prominent steep band of velocities in the center. 
On the receding side, the velocities stay more or less constant once they reach the plateau.  
The second component also shows a similar steep band of velocities as the first component, but it is less pronounced and much wider. 
The orange and red points between 0\arcsec < r < 500\arcsec, 0 km s$^{-1}$ < v < 100 km $^{-1}$ and 100\arcsec < d < 200\arcsec correspond to the arm of zero velocity visible in Fig. \ref{fig:second_comp}. 
To the left of the center, there is an almost flat band of negative velocities, this 
is the zone of approaching velocities on the eastern side of the bulge in 
Fig. \ref{fig:second_comp}. 
\begin{figure}
\resizebox{\hsize}{!}{\includegraphics{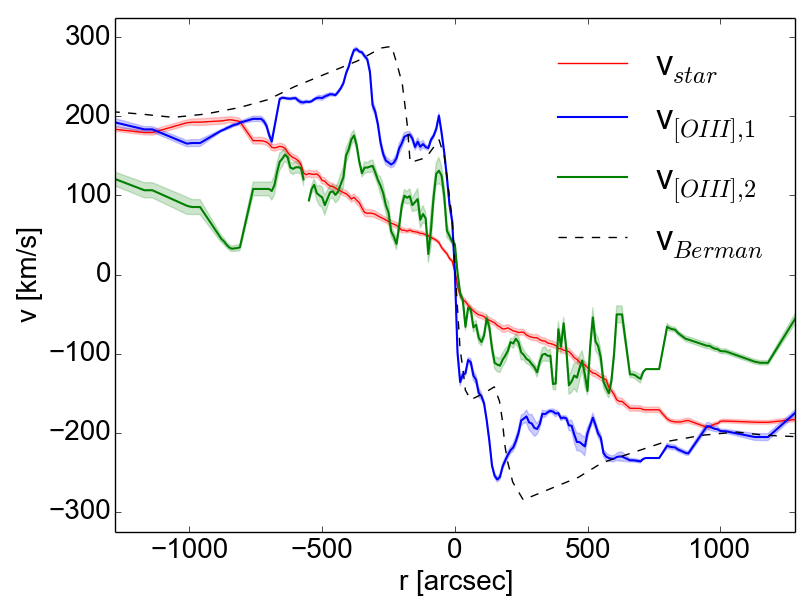}}
\caption[Major axis cut through the velocity maps]{Cut through the velocity maps along the disk major axis within an aperture of 40\arcsec. The red line is a cut through the stellar velocity (Fig. \ref{fig:v_star}), the blue one is a cut through the first velocity component (Fig. \ref{fig:first_comp}), the green 
one the second velocity component (Fig. \ref{fig:second_comp}). The black dashed line is the velocity for the triaxial bar model by \citet{Berman01}.}
\label{fig:cut_major_axis}
\end{figure}

The position velocity diagram of the first gas component agrees very well with the overall shape 
of the model of \citet{Berman01}, who simulates a triaxial bulge in M31: the velocity reaches a peak, then stays constant and rises to 
a second peak. The comparison plot of his data to cuts through our maps is shown in Fig. \ref{fig:cut_major_axis}. However, our position-velocity diagrams are asymmetric: for the approaching side of M31 the peak is closer to the center than for 
the other side. The shape of the PV-diagram could also be compatible with the superposition of other structures, like rings and discs, but this discussion goes beyond the scope of this paper. 
The overall shape of the position velocity diagram is very similar to what is seen in simulations 
 \citep{Athanassoula99} and observations of barred galaxies \citep{Bureau99, Merrifield99}.

The pattern that is visible in the flux maps in [OIII] and H$\beta$ is similar to the one seen by \citet{Jacoby85} in a H$\alpha$+[NII] filter and in [OIII], by \citet{Boulesteix87} in [NII] and by \citet{Ciardullo88} in H$\alpha$+[NII].
As seen in section \ref{sec:flux}, the H$\beta$ and [OIII] maps have the overall appearance of 
a spiral structure, with an incomplete elliptical ring 
inside, which is oriented roughly along the minor axis, with a ``spoke'' along the ring's short axis. 
The 100\arcsec long ``spoke'' along the disk major axis could be an inner disk that is projected due to M31's inclination and orientation into the elongated shape we see \citep{Jacoby85}. 
The inner spiral pattern 
seems to be tipped to a lower inclination with respect to the outer part, which according to \citet{Jacoby85} can be caused by a non-axisymmetry, like a bar, 
and cannot be explained by axisymmetric features alone.

While the ring structure lies in the region where \citetalias{Blana16a} find two inner Lindblad resonances, it is hard to deproject the structures we see in our images. A detailed analysis of the structures is postponed to 
      a later paper, where gas will be taken into account in the dynamical model.

      The fact that the incomplete ring is aligned along the minor axis, almost perpendicular to the position of 
the assumed bar, led \citet{Block06} to conclude that this ring 
is not caused by a bar, but by the collision of M31 with M32, which also caused a ``split'' of the so-called 10-kpc ring, which is a structure appearing further out 
in the gas. 

While the collision model is in better agreement with the ring structure, especially with the fact that the rings are 
off-center, it has lower flux inside the ring, whereas in our flux maps, as well as the ones by \citet{Jacoby85}, \citet{Boulesteix87} and 
\citet{Ciardullo88}, there is flux present there. 
We therefore think that the bar is a more likely explanation for the rings. However, as mentioned above, to fully reproduce the visible gas rings, we need to include gas into our bar models, which 
will be done in the future.

\subsection{Bar or triaxial bulge?}
{The first models that tried to explain the triaxiality in M31
\citep{Stark77, Stark94, Berman01, Berman02} called their structures a
``triaxial bulge'' instead of a  ``bar''.  These two concepts are fundamentally
different: A bar is a rotating structure that results from an instability in
the disk, while a triaxial bulge is a consequence of an anisotropy in velocity
dispersion, generated, e.g.\ by merging processes.  However, as already stated
in section \ref{sec:Bar_intro}, the models in the aforementioned papers
actually describe rapidly rotating systems, which in our understanding are
``bars''. \\
After buckling, most observed bars consist of a three-dimensional inner part of
the bar, the B/P-bulge, and a flat outer part (see e.g.
\citealp{Athanassoula05} or \citealp{Martinez-Valpuesta06}). In this
nomenclature, the ``triaxial bulge'' of the models by \citet{Berman01} and
\citet{Berman02} is a B/P-bulge. \\ Nevertheless, it is still interesting to
see if our findings are also consistent with a completely non-rotating
structure. \\ 
Firstly, we attempt to quantify the level of cylindrical rotation in M31, as
cylindrical rotation is generally attributed to the presence of a bar
\cite{Bureau2006}.  Based on \citet{Saha2013}, \citet{Molaeinezhad16}
(hereafter M16) define the parameter $m_{cyl}$ = $m_{avg} + 1$ where $m_{avg}$
is a measure of the change in rotational velocity in the direction
perpendicular to the major axis.  $m_{avg}$ averages over the slopes of the
decrease in rotational velocity within the bulge region at various distances to
the minor axis. M16 first test the discriminatory power of this quantity on
data from the Giraffe Inner Bulge Survey \citet{Zoccali2014} of the Milky Way
and on a barred $N$-body model of the Milky Way by
\citet{Martinez-Valpuesta2011}, and then apply it to 12 further disk galaxies
with inclinations ranging from 70\degr to 90\degr. They find that the B/P
bulges in their sample as well as the Milky Way bulge show levels of
cylindrical rotation of $m_{cyl} \simeq 0.7$.\\
As M16 we derive rotation curves parallel and at varying distances to the major
axis of M31 first and then turn these into one-sided rotation curves parallel
to the minor axis at varying distances to the minor axis. We perform this
analysis on the south eastern half of M31 as it is this half where the disk is
located on the far side, such that the bulge kinematics are not affected by
dust screening of the disk. For the extent of the bulge along the major axis
and along the minor axis we adopt values of 460\arcsec \ and 270\arcsec \
respectively from the bulge to disk decomposition in \citet{Kormendy99} and use
the photometric disk position angle of 38$\degr$ \citep{deVaucouleurs58}. The
resulting plot is shown in Fig.\ \ref{fig:Cylindrical_rot}.  The parameter
$m_{cyl}$ for the south-eastern side is 0.72. Changing the exact values of the
position angle and also varying the radii for the bulge in major axis direction
and in minor axis direction, and also the area of exclusion around the major
axis (we use 50\arcsec) have a minor effect on $m_{cyl}$. Tests show that for
reasonable parameter ranges the value never becomes smaller than 0.68.  Hence
the level of cylindrical rotation --- as quantified by this method --- seems to
be consistent to what is reported in \citet{Molaeinezhad16} for their B/P
bulges.\\ 
\begin{figure}
 \resizebox{\hsize}{!}{\includegraphics{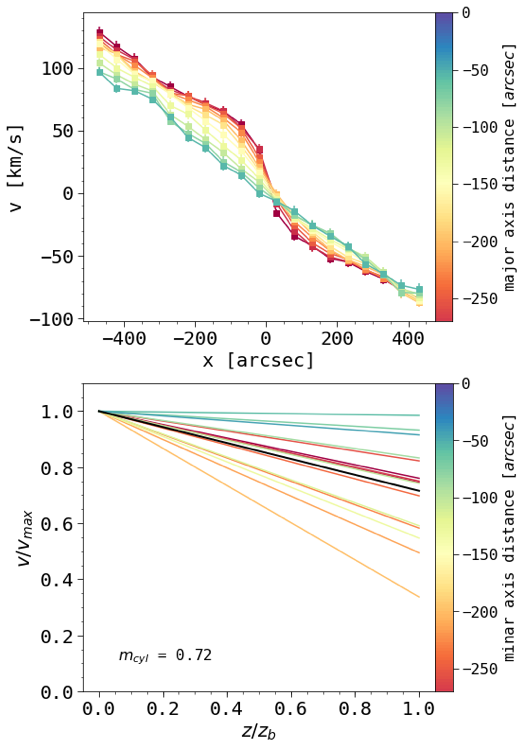}}
 \caption[Measurement of cylindrical rotation]{Measurement of the cylindrical
rotation of the south-east side of M31 according to the method by
\citet{Molaeinezhad16}.Top: The stellar line-of-sight velocity parallel to the
major axis of galaxies at different heights from the disk plane. Bottom:
The Line-of-sight velocity gradient parallel the minor axis at different
distances from the minor axis. Profiles are grey-scale-coded according to the
distance from the minor axis. The solid black profile represents the average
value $m_{avg}$. The cylindrical rotation is then $m_{cyl}$ = $m_{avg} + 1$ as
a quantity to express the level of cylindrical rotation. }
 \label{fig:Cylindrical_rot}
\end{figure}
Secondly, as described in \citet{Athanassoula06} and \citetalias{Blana16a}, the
spurs in the photometry outside the bulge argue for a planar bar, the ``thin''
or ``long bar'' component described in both papers. Although both papers give
different lengths for the thin-bar components, the fact still remains that such
a component is required to explain the photometric appearance of the galaxy. \\
Thirdly, the outer 10\,kpc ring of M31 is associated with the outer Lindblad
resonance of the bar by \citetalias{Blana16a}.  This structure is only possible
in a rotating barred potential, as there are no characteristic resonance radii
in non-rotating triaxial potentials. In a system with a triaxial bulge, this
ring would be an ad-hoc feature which is difficult to explain. \\ 
Finally, the correlation between v and h3 (see Fig. \ref{fig:h3correlation}) is
a clear signature for a bar and not a triaxial bulge.}

\subsection{Explanations for multiple components}
Several scenarios have been suggested by various authors for the origin of the multiple gas components.
When gas is subject to a non-axisymmetric potential like a bar, it leads to the formation of gas streams and rings, see 
e.g. \citet{Kim12} for numeric models. The line-of-sight passes through several of these streams, 
therefore leading to multiple peaks in the gas lines. A detailed comparison will only be possible once 
gas is included into the dynamical model, which will be done in a future paper.

There is also an alternative explanation for such streams and rings in M31, which is that they were created by 
the collision of M31 with its small companion galaxy M32 \citep{Block06}. 
A galaxy with similar inclination as M31, which has also a B/P bulge, is NGC 2683 \citep{KuziodeNaray09}. Investigating the H$\alpha$ velocity field, they see S-shaped twists 
and argue for the presence of a bar at a position angle of 5$\degr$ higher than the disk position angle.
Differences in their PV-diagram and the ones for M31 led \citet{Melchior11} to conclude that the interpretation of \citet{Athanassoula06} in terms of a  
bar in M31 is wrong and that the ring structures are only due to the collision of M31 with M32 suggested by \citet{Block06}.
However, the bar in NGC 2683 is only at an angle of 5$\degr$ to the disk major axis, so it is seen quite side-on, whereas 
the bar in M31 is at an angle of 17$\degr$ to the disk major axis and it is seen more end-on, 
which can explain the differences according to \citet{KuziodeNaray09}.

\citet{Melchior11} also measure molecular emission lines in CO with two peaks, in \citet{Melchior16} sometimes even three. While the molecular gas is not expected to follow the ionized gas, 
it is still interesting that they also observe a line splitting.
\citet{Melchior11} try to explain the double CO components with a tilted ring model coming from the collision model by \citet{Block06}. An small disk with inclination 
of 43$\degr$, i.e. more face-on than the large-scale disk of M31, is surrounded by a ring. 
The velocity profiles extracted from the simulated velocity fields by 
\citet{Melchior11} consist of a very broad component, which is blueshifted and a narrow one, which is redshifted, 
the blueshifted part comes from the inner disk and the redshifted part from the ring. Our [OIII] measurements taken at the same positions look quite different, both components roughly have the same width, so the scenario of \citet{Melchior11} does not predict the actual shape of our measured [OIII] spectrum.

Another scenario for multiple components is that large-scale warps in the outer galaxy disk project
lower velocities from the outer regions into the inner disk. In the neutral HI gas, \citet{Chemin09} find up to five different gas components in M31. The velocity maps for all these 
components look very similar and have basically the same regular kinematic pattern. A component with a steep slope in velocity in the center 
is the main HI component, while components with flatter slopes are due to slower gas in the outer HI disk that is projected into the central 
areas due to warps in the HI disk. However \citet{Athanassoula06}  find that 
the different slopes can be explained by the streaming gas motions caused by the bar. Comparing the PV-diagrams of \citet{Chemin09} with our own, we find that their ``main HI component'' agrees with the 
first gas component we measure, while the ``zero velocity spiral arm'' we measure in the second component agrees with warped components in HI. However, it should be kept in mind that the 
spatial resolution in \citet{Chemin09} in the center is less than our own.\\

\subsection{Ionization mechanisms of the gas}
\begin{figure}
 \resizebox{\hsize}{!}{\includegraphics{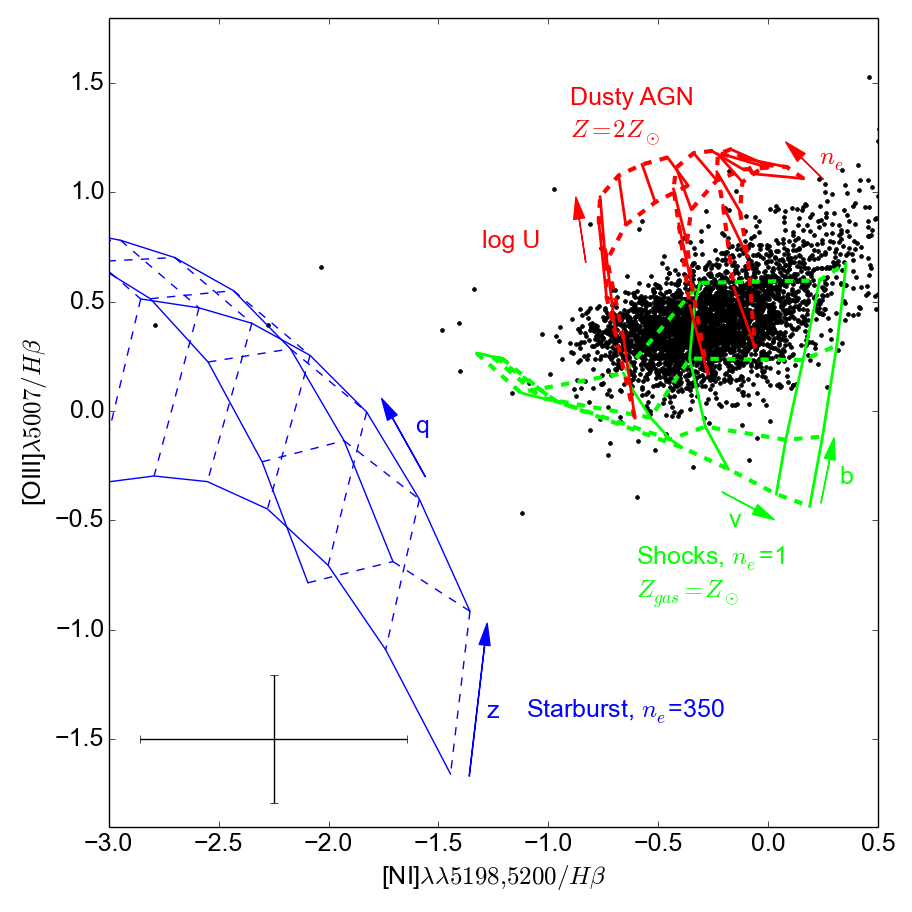}}
 \caption{The diagnostic diagram for the first gas component. The plot is equivalent to the right panel of Fig.1 in \citet{Sarzi10}. The black points are the values from our data. The median error values of the errorbars are shown in the lower 
 left corner. Overplotted are lines showing the predictions for gas that is photoionised by a central AGN, 
 by O-stars or by shocks, respectively. The AGN grids are from the dusty, radiation pressure-dominated models of \citet{Groves04} and adopt three values for the index $\alpha$ of the power-law AGN 
 continuum $f_{\nu} \propto \nu^{\alpha}$, with $\alpha$= -1.7, -1.4, -1.2 from left to right. In each AGN model grid, the solid lines trace the dimensionless ionization parameter log U (defined as log q/c), which 
 increases with the [OIII]/H$\beta$ ratio from $log U = -3.0, -2.6, -2.3, -2.0, -1.6, -1.3, -1.0$, whereas the dashed lines show the adopted values for the electron density of $N_e=10^2$ and $10^4$ cm$^{-3}$, 
 with smaller values of the [NI]/H$\beta$ ratio corresponding to larger $N_e$ values. The starburst grids are from \citet{Dopita00}, and assume a gas density of $N_e=350$ cm$^{-3}$ and use a 
 spectral energy distribution obtained from models from Starburst99 \citep{Leitherer99} for an instantaneous star formation episode. The grids assume a range of metallicity $Z$ for both stars and gas in the starburst, shown by the solid lines for $Z$ = 0.2, 0.4, 1.0, 2.0 $Z_{\odot}$, and different 
 values for the ionization parameter $q$, shown by the dashed lines for $q = 0.5, 1, 2, 4, 8, 15, 30 \times 10^7 $cm s$^{-1}$. The shock grids, without precursor HII region, are from \citet{Allen08} and assume an electron density of $N_e$=1$cm^{-3}$ and a solar value for the gas 
 metallicity. The solid lines show models with increasing shock velocity $v$ = 150, 200, 300, 500, 750, 1000 km s$^{-1}$, and the dashed lines models with magnetic parameter b= 0.5, 1.0, 2.0, 4.0. The grids have been obtained using the program \texttt{ITERA} \citep{Groves10}. We assume solar metallicity for the 
 shocks and twice solar metallicity for the dusty AGN grids, because we obtain these values in the bulge in our stellar population measurements, which will be presented in \citet{Opitsch16}.}
 \label{fig:BPT_diagram_1}
\end{figure}

\begin{figure}
 \resizebox{\hsize}{!}{\includegraphics{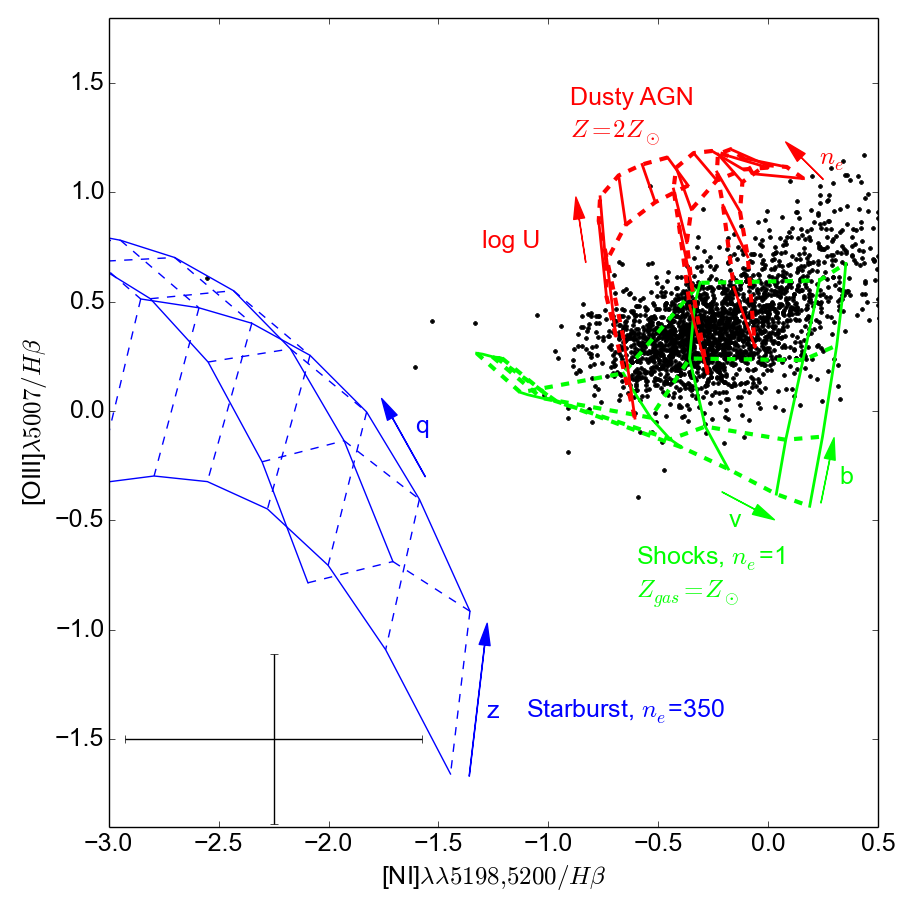}}
 \caption{The diagnostic diagram for the second component. The plot is equivalent Fig. \ref{fig:BPT_diagram_1}, with the grids being 
 the same.}
\label{fig:BPT_diagram_2}
\end{figure}

In order to investigate which mechanisms are responsible for ionizing the gas, diagnostic diagrams are used, which compare the ratios of line fluxes of different 
emission lines. The most widely utilized of these diagrams compares $[OIII]\lambda5007/H\beta$ to $[OI]\lambda6300/H\alpha$ \citep{Veilleux87}.
Since we don't have H$\alpha$ or $[OI]\lambda6300$ in our observed wavelength range, we cannot use this standard diagnostic diagram.
\citet{Sarzi10} devised alternative diagnostic diagrams for the \texttt{SAURON} spectrograph \citep{Bacon01}, which has a similar wavelength range as \texttt{VIRUS-W}. 
This diagram compares $[OIII]\lambda5007/H\beta$ to $[NI]\lambda\lambda5198,5200/H\beta$. The [NI] lines are usually present in partially ionized regions in gaseous 
nebulae, which are photo-ionized by a spectrum containing high-energy photons, but they are absent in HII regions, where $H\beta$ and $[OIII]\lambda5007$ arise.
The Sarzi diagram for the first [OIII] component is shown in Fig. \ref{fig:BPT_diagram_1}, the one 
for the second component in Fig. \ref{fig:BPT_diagram_2}. There is not much difference between the 
two diagrams, the datapoints occupy similar regions in both.

The ionization does not happen via starbursts, instead most of the points lie in the 
region where the ionization is due to shocks in the gas. That shocks are the primary source for ionization is supported by the fact that the emission seems to be strengthened along the lines of the velocity discontinuities. These shocks could be triggered by the bar.
About half the points, however, also lie in between the Shock region and the AGN region, which is occupied by Seyfert 
galaxies in \citet{Sarzi10}, but since the AGN in M31 is very weak \citep{delBurgo00}, excitation by the AGN is unlikely. These datapoints lying between the regions can be partially explained by the fact that the errors of the datapoints are large. Furthermore, the diagnostic diagram used here does not present as clear separations
    as the usual [OI]/H$\alpha$ vs. [OIII]/H$\beta$ diagram, compare \citet{Sarzi06}. Contamination by planetary nebulae is also not an issue, we checked the 166 bins that are affected by PNe. They lie in the upper parts of the shock grid and  between the shock and the AGN grid, there is no systematic offset of the PNe positions
    with respect to 
    the rest of our spectra.

\section{Conclusions}
\label{sec:Conclusions}
\subsection{Main results}
For several decades, the question has been posed if M31 is a barred galaxy or 
not (\citealp{Lindblad56, Stark77, Stark94, Athanassoula06};~\citetalias{Blana16a}).
The high inclination of M31 made it difficult for a long time to see the bar directly in optical photometry. 
Twists in the inner isophotes were the first hints for its existence \citep{Lindblad56, Stark77}, while more recently,
the boxy appearance of the isophotes in infrared images was used to constrain its orientation and length 
(\citealp{Beaton07}; ~\citetalias{Blana16a}). However, a kinematic confirmation 
of the bar has been missing so far. \\
In this paper, we have presented results from observations of M31 with the optical integral field unit spectrograph
\texttt{VIRUS-W}.  We derive the kinematic properties of the stars and two different ionized gas components and measure the gas fluxes. We find hints in the kinematics that M31 contains a bar.
Our main results are:

\begin{enumerate}
\item Cuts through the stellar velocity 
field reveal a plateau at moderate radii, which is a signature that can be created by a bar that is seen at an intermediate angle \citep{Bureau05}.
The line of zero velocity does not coincide with the photometric minor axis and shows a twist towards the edge of the bulge.
\item The stellar velocity dispersion field shows a drop in the center, with two maxima aligned along the minor axis. 
While this central drop can be caused by inner disky structures or dust in the center 
\citep{Falcon-Barroso06} or the classical bulge \citepalias{Blana16a}, it can also be due to the bar, since the stellar orbits 
that make up the bar become more circular in the center \citep{Bureau05}. There are also two inclined plateaus at intermediate 
radii in the $\sigma$ profile, a  behavior that is also often seen in barred potentials \citep{Bureau05}.
\item The higher Gauss-Hermite moment $h3$ is anti-correlated with the velocity $v$ in the disk regions and the very center, 
which is the expected behavior for a disk component \citep{Bender94}. 
In the majority of the bulge region, $h3$ is correlated with the velocity $v$, which can 
be achieved by elongated motions which occur along the bar direction \citep{Bureau05}.
\item The gas kinematics, measured using the [OIII]$\lambda$5007 line, is more complicated than the stellar kinematics. Many spectra exhibit two separate peaks, resulting in two kinematically distinct components. 
One of the two components has faster velocities than the other one.
The fast component has disk-like rotation with a very steep gradient in the center.
The line of zero velocity is S-shaped, again pointing to a bar. The overall shape of the rotation curve 
of the first component is qualitatively in agreement with gas in a barred dynamical model of M31 by \citet{Berman01}.
\item The position velocity diagrams of the gas components look similar to what is expected from 
simulations \citep{Athanassoula06}, with arms of steep and flatter slopes. The overall shape of the position-velocity diagram agrees with observations in HI by 
\citet{Chemin09}. They measure up to 5 different components, the most luminous of which is the main HI disk, which 
coincides with the steep slope in the position-velocity diagram.
The other components belong to the branches with lower slope in 
the position-velocity diagrams. According to \citet{Chemin09}, they are low velocities from the outer regions 
of the HI disk, which have been projected to the center due to warps.
However, they can also agree with the lower velocity branches that exist in a barred 
potential \citep{Athanassoula06}. 

\item When looking at the morphology of the gas, we see a spiral pattern, similar to what is seen by \citet{Jacoby85} in H$\alpha$+[NII] and [OIII], \citet{Boulesteix87} 
in [NII] and \citet{Ciardullo88} in H$\alpha$+[NII]. This spiral pattern has a lower inclination than the disk, which means it could have been tilted by a non-axisymmetry, like a bar \citep{Jacoby85}.

\item \citet{Block06} find that the morphology of M31 in the far infrared is not caused by a bar, but is instead the 
result of 
a density wave caused by the collision of M32 with M31. Comparing their gas morphology with our measured one, the ring somewhat corresponds to what we see, but there 
is no high emission inside the ring, which we observe in [OIII]. Building on this model, \citet{Melchior11} propose a scenario of a tilted ring in the center over 
another rotating disk, to explain line splittings they measure in CO observations. Their model predicts that the component in the ring has a very 
narrow line, while the disk leads to a very broad line, which is not what we see, our two components in that region have the same width. 
We therefore think that this model does not predict our observations.

\item From diagnostic diagrams for the ionization of the gas, we find that the gas is mostly ionized by shocks and not by starbursts, which is consistent with 
the low star formation rate of M31 \citep{Davidge12}. This is also in agreement by the filamentary appearance of the gas morphology, which could 
be either due to shocks or supernovae of type Ia \citep{Jacoby85}. 

\end{enumerate}
To summarize, there are several hints in the kinematics of the stars and the ionized gas and its morphology for a bar in M31.

\subsection{Future work} 
From the analysis of the absorption line strengths, we measured the age, metallicity and $\alpha$/Fe-overabundance of the stellar populations. These results will 
be presented in an accompanying paper \citep{Opitsch16}. These measurements,  combined with the kinematics presented in this paper, 
are used as a basis for a made-to-measure dynamical model \citep{Blana16b}. This model will allow us to truly test the bar hypothesis and to check if our observations 
are caused by the bar, especially once gas is taken into account. They will also provide new estimates on the number of microlensing events caused by MACHOs or self-lensing, which will be used in future microlensing surveys towards M31.
We also plan to observe further pointings along the bar major axis, to calculate the pattern speed of the bar. 

\bibliographystyle{aa}
\bibliography{Opitsch_et_al}

\section{Appendix}
\subsection{Comparison with data from \citetalias{Saglia10}}
\label{sec:Saglia}
\begin{figure}
\begin{center}
\resizebox{\hsize}{!}{\includegraphics{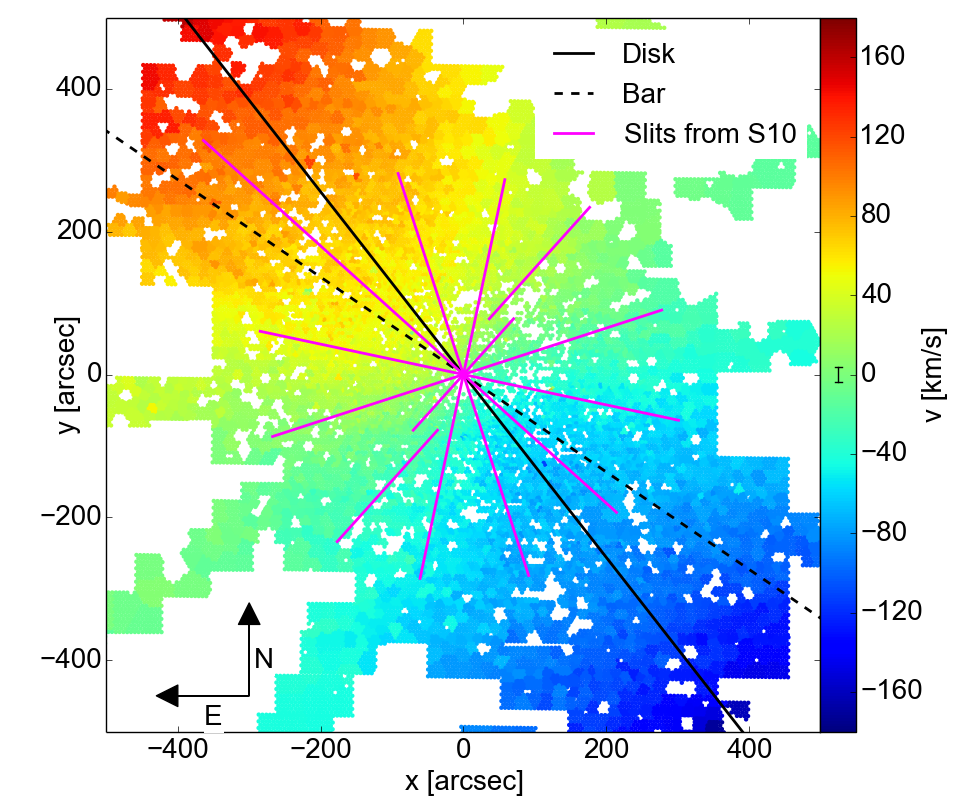}}
\caption[Velocity map with slits from \citetalias{Saglia10}]{Velocity map with disk major axis (PA=38$\degr$, solid black line), the bar major axis (PA=55.7$\degr$, dashed black 
line) and the slit positions by \citetalias{Saglia10} (magenta). They are PA=48$\degr$ (the bulge major axis), PA=78$\degr$, PA=108$\degr$, PA=138$\degr$ (the bulge minor 
axis), PA=108$\degr$ and PA=18$\degr$, all angles measured east from north. The line in the colorbar is the median errorbar of the velocities.}
\label{fig:Saglia_coverage}
\end{center}
\end{figure}

\begin{figure}
 \resizebox{\hsize}{!}{\includegraphics{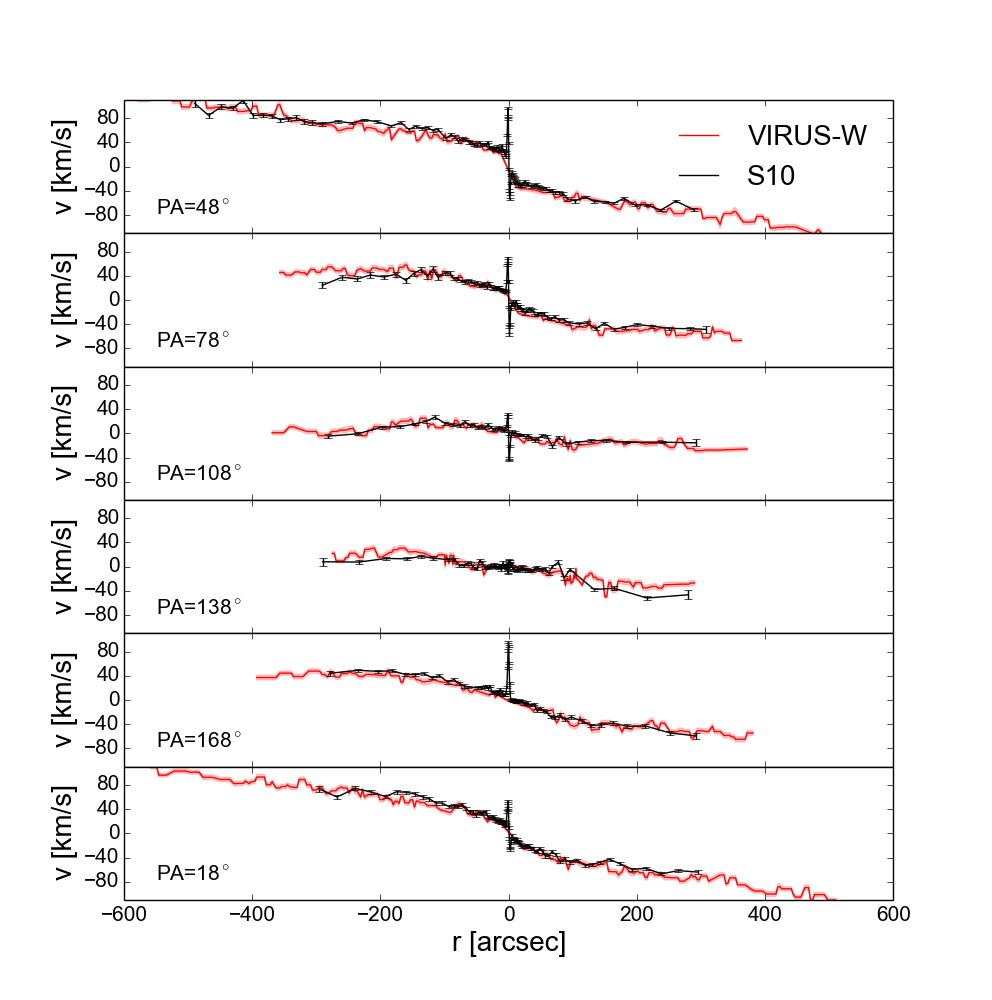}}
 \caption[Comparison of our velocities with the ones from \citetalias{Saglia10}]{Comparison of our velocities with the ones from \citetalias{Saglia10}. Black are the velocities measured 
 by \citetalias{Saglia10}, red are cuts through our velocity maps along the same directions.}
 \label{fig:comparison_Saglia_v}
\end{figure}

We compare the stellar and gas kinematics measured with \texttt{VIRUS-W} with the ones from \citetalias{Saglia10}. \citetalias{Saglia10} observed M31 with a longslit spectrograph along 6 directions, the slit positions are plotted 
in Fig. \ref{fig:Saglia_coverage}. 
We make cuts through our own stellar velocity maps along the slit directions of \citetalias{Saglia10} and compare them with their data in Fig. \ref{fig:comparison_Saglia_v}.
Our velocities agree within 5$\%$ with the ones by \citetalias{Saglia10}, except for the minor axis on the near side (positive radii in Fig. \ref{fig:comparison_Saglia_v}), where the deviation 
is about 30$\%$.

\begin{figure}
\resizebox{\hsize}{!}{\includegraphics{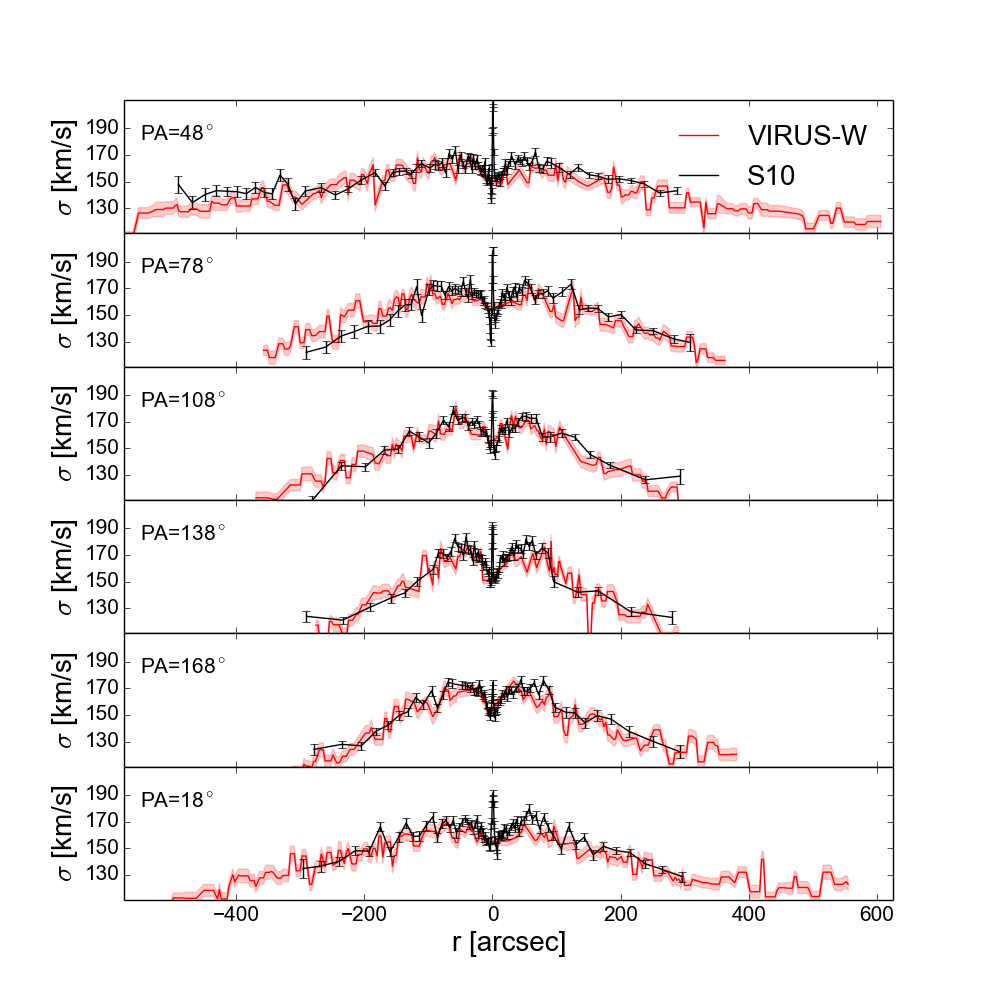}}
 \caption[Comparison of our velocity dispersions to the ones from \citetalias{Saglia10}]{Comparison of our velocity dispersions with the ones from \citetalias{Saglia10}. Black are the values measured 
 by \citetalias{Saglia10}, red are cuts through our velocity dispersion maps along the same directions.}
 \label{fig:comparison_Saglia_sigma}
\end{figure}

The comparison plot for the velocity dispersion is shown in Fig. \ref{fig:comparison_Saglia_sigma}. 
Overall, our data agree well with the measurements of \citetalias{Saglia10}, but we do not reproduce the central spike in $\sigma$ that is caused by the supermassive
black hole, because we lack resolution in the very central regions. The overall agreement is within 4$\%$.

\begin{figure} 
\resizebox{\hsize}{!}{\includegraphics{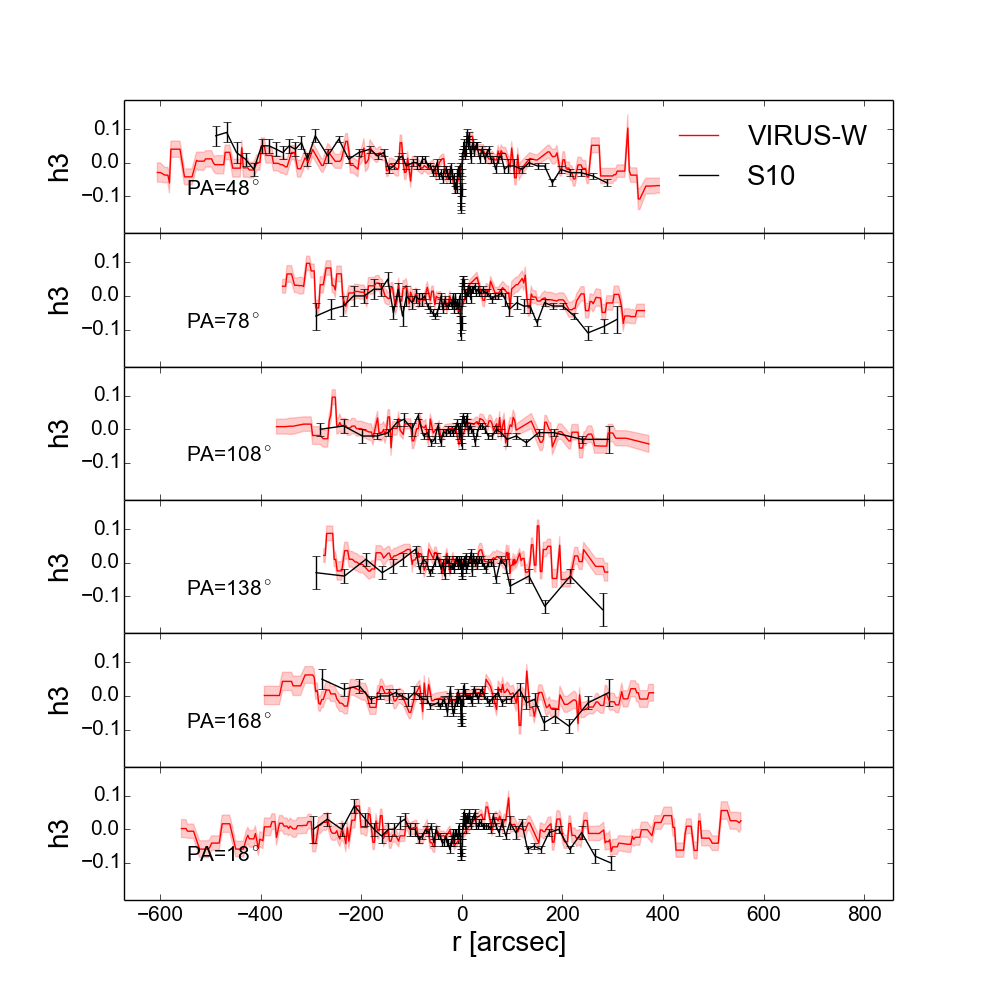}}
 \caption[Comparison of our h3 measurements to the ones from \citetalias{Saglia10}]{Comparison of the measured $h3$ values with the ones from \citetalias{Saglia10}. Black are the values measured 
 by \citetalias{Saglia10}, red are cuts through our maps along the same directions.}
 \label{fig:h3_Saglia}
\end{figure}

\begin{figure} 
 \resizebox{\hsize}{!}{\includegraphics{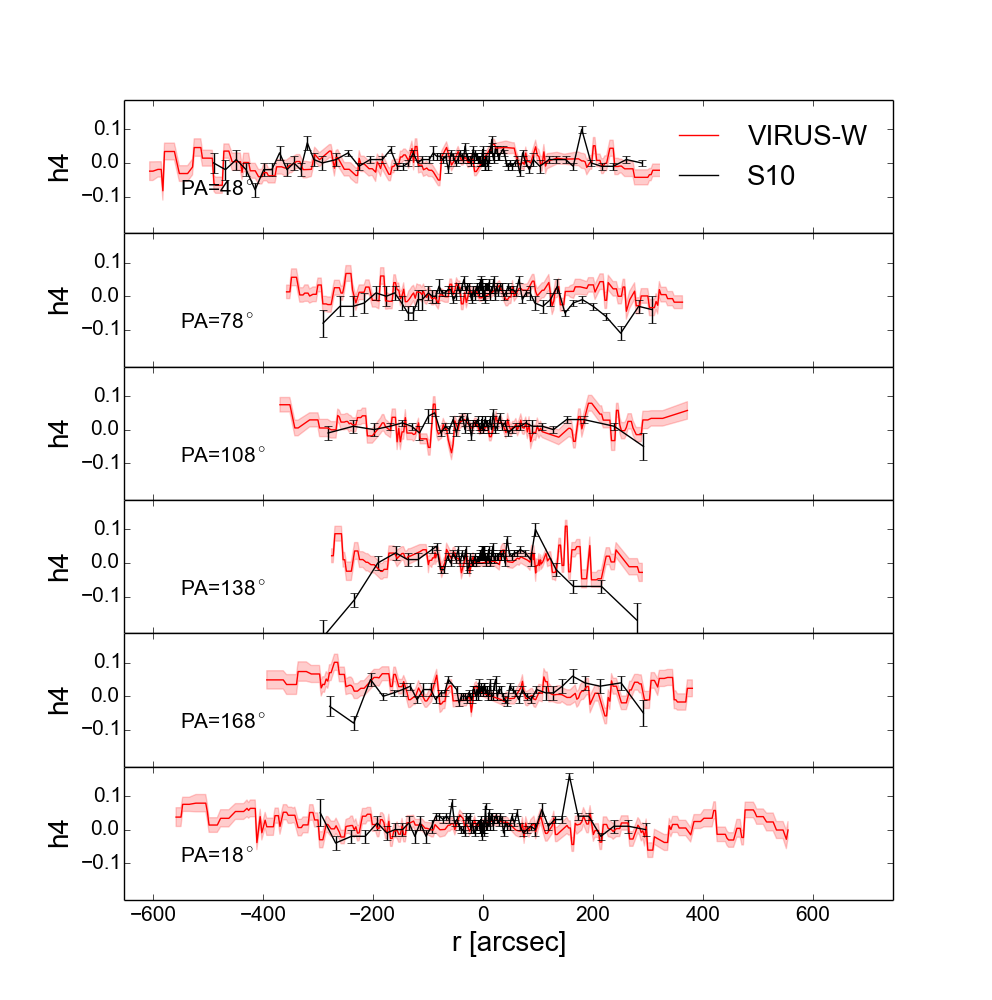}}
 \caption[Comparison of our h4 measurements to the ones from \citetalias{Saglia10}]{Comparison of the measured $h4$ values with the ones from \citetalias{Saglia10}. Black are the values measured 
 by \citetalias{Saglia10}, red are cuts through maps along the same directions.}
 \label{fig:h4_Saglia}
\end{figure}

The comparison for $h3$ can be seen in Fig. \ref{fig:h3_Saglia}.
The standard deviation of the difference between our values and the ones by \citetalias{Saglia10} is 0.020 - 0.027, which 
agrees with the root mean square value of the error of the \texttt{VIRUS-W} values.\\
In Fig. \ref{fig:h4_Saglia}, the corresponding cuts for $h4$ are shown.
Again, we find good agreement, the standard deviation of the difference between our values and the ones by \citetalias{Saglia10} is 0.02, which 
agrees with the root mean square of the error of the \texttt{VIRUS-W} values.\\

\begin{figure}
\resizebox{\hsize}{!}{\includegraphics{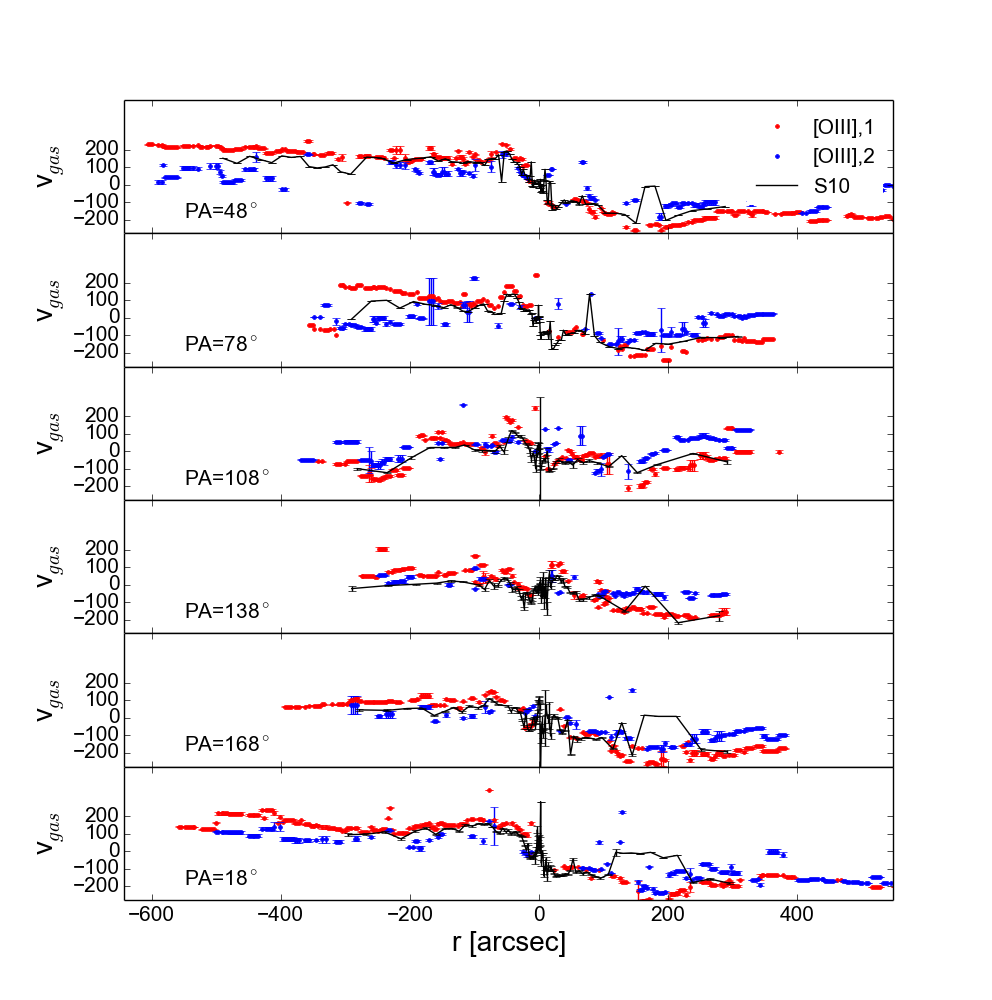}}
\caption[Comparison of cuts with data from Saglia et al., 2010]{Comparison of our gas velocities with the ones from \citetalias{Saglia10}. Black are the velocities measured 
 by \citetalias{Saglia10}, red are cuts through the first gas component from Fig. \ref{fig:first_comp} and blue the cuts through the second component from Fig. \ref{fig:second_comp}.}
 \label{fig:Saglia_v_gas}
\end{figure}

We also compare our gas kinematics to the data from \citetalias{Saglia10}.
The comparison plot is shown in Fig. \ref{fig:Saglia_v_gas}. Generally, the first component agrees with the values from 
\citetalias{Saglia10}. For large radii, the velocity from \citetalias{Saglia10} often lies between both 
velocity components. This is somewhat expected, because \citetalias{Saglia10} do not 
resolve the two components, so they measure a broader Gaussian, which incorporates both
thin Gaussians from the two components. This broader Gaussian has of course a peak value that 
is in between the two thin ones, which is then adopted as the velocity.

The gas velocity dispersions of \citetalias{Saglia10} are affected by the relatively poor spectral resolution ($\approx$ 69 km s$^{-1}$) and tend to be systematically larger than our values.
Overall, our data agree very well with \citetalias{Saglia10}.

\newpage

\begin{table}
\caption{Lines used for wavelength calibration}
\label{tab:HgNe}
\begin{tabular}{l r}
 \hline
 \hline
Wavelength \AA & Ionization state  \\
\hline
4837.314     &Ne I  \\
4863.081      & Ne I \\
4884.917   & Ne I \\
4916.068   & Hg I \\
4957.036   & Ne I \\ 
5005.159   & Ne I \\
5025.600    	& Hg II \\
5031.348     & Ne I \\
5037.751    & Ne I \\
5074.201    & Ne I \\
5080.383    & Ne I \\
5113.672    & Ne I \\
5116.503    & Ne I \\
5122.257    & Ne I \\
5144.938    & Ne I \\
5151.961    & Ne I \\
5188.612    & Ne I \\ 
5203.896    & Ne I \\
5208.865    & Ne I \\
5222.352    & Ne I \\
5234.027    & Ne I \\
5298.190    & Ne VI \\
5330.778    & Ne I \\
5341.094    & Ne I \\
5343.283    & Ne I \\
5400.562    & Ne I \\
5433.651    & Ne I \\
\hline
\end{tabular}
\end{table}

\begin{table}
\caption{Kinematic standard stars, taken from the \texttt{ELODIE} catalog \citep{Prugniel07}}
\label{tab:kinematic_standards}
\centering
\begin{tabular}{r l r r}
 \hline \hline 
  HD(...) & Type  & $RA $ & $DEC$  \\
          &       & J2000.0 & J2000.0 \\
  \hline
 5516 & G8III-IV & 00:57:02.0 &   +23:24:04.0   \\
6203 & K0III-IV & 01:03:02.5 &   -04:50:11.8    \\
7010 & K0IV &     01:11:37.6 &   +60:30:22.6    \\
10380 & K3-IIIb & 01:41:25.9 &   +05:29:15.4   \\
12929 & K2IIIab & 02:07:10.4 &   +23:27:44.7   \\
19476 & K0III & 03:09:16.0 &   +44:50:51.0         \\
20893 & K3III & 03:22:45.2 &   +20:44:31.6        \\
27348 & G8III & 04:20:12.0 &   +34:33:33.0          \\
30834 & K3III & 04:52:25.0 &   +36:41:53.0     \\
35369 & G8III & 05:23:47.0 &   -07:48:37.0     \\
37160 & G8III-IV &05:36:44.0 &   +09:17:35.0   \\
38309 & F0III:n & 05:42:23.4 &   +03:59:19.0      \\
39003 & K0III & 05:51:16.0 &   +39:08:52.0     \\
39118 & G8III+... & 05:47:53.9 &   +02:00:41.4     \\
39833 & G0III & 05:52:29.1 &   -00:30:53.9        \\
42787 & M2III & 06:12:59.5 &   +06:00:58.5       \\
43039 & G8IIIvar &06:15:22.0 &   +29:29:55.0   \\
45415 & G9III & 06:27:20.4 &   +02:54:29.9         \\
46377 & K4III & 06:30:30.9 &   +01:18:42.5       \\
46784 & M0III & 06:32:48.6 &   +05:33:13.8        \\
48433 & K1III & 06:43:48.0 &   +13:13:56.0          \\
54079 & K0III: &07:07:49.4 &   +07:28:16.3     \\
54489 & G9III & 07:09:07.7 &   +02:15:11.1     \\
58207 & G9III+...&  07:25:43.0 &   +27:47:53.0   \\
58923 & F0III & 07:28:02.0 &   +06:56:31.0     \\
61935 & K0III & 07:41:14.0 &   -09:33:03.0     \\
62345 & G8III & 07:44:26.0 &   +24:23:53.0     \\
62437 & F0III & 07:41:31.2 &   +02:31:33.7        \\
72561 & G5III & 08:31:05.3 &   +04:55:43.4     \\
76294 & G8III-IV &08:55:23.0 &   +05:56:43.0   \\
81192 & G7III & 09:24:46.0 &   +19:47:17.0     \\
94672 & F2III & 10:53:08.0 &   +01:00:14.3     \\
104979 & G8III & 12:05:13.0 &   +08:43:56.0    \\
120136 & F7V & 13:47:16.0 &   +17:27:24.0      \\
122563 & F8IV & 14:02:32.0 &   +09:41:14.0     \\
137759 & K2III & 15:24:55.0 &   +58:57:57.0    \\
169959 & A0III & 18:26:52.9 &   +06:25:24.5    \\
171802 & F5III & 18:36:27.8 &   +09:07:21.1    \\
215648 & F6III-IV & 22:46:41.6 &   +12:10:22.4    \\ 
\hline
\end{tabular}
\end{table}

\begin{table}
\caption{Photometric standard stars}
\label{tab:photometric_standards} 
\begin{tabular}{l l c r c}
\hline
\hline
Name & Type  & $RA $ & $DEC$  & Reference\\
       &       & J2000.0 & J2000.0 &  \\
  \hline
BD+284211 & Op & 21:51:11.1 &  +28:51:52.0 & 1\\
Feige 66 & sdO & 12:37:23.6 &  +25:04:00.0 & 1\\
Feige 110 & DOp& 23:19:58.4 &  -05:09:56.0 & 1\\
HD 84937 & sdF5& 09:48:55.9 &  +13:44:46.1 & 2\\
\hline
\end{tabular}
\tablebib{ (1)~\citet{Oke90}, (2) \citet{LeBorgne03}}
\end{table}

\begin{landscape}
  \begin{table}
\caption{Kinematic data of the stars. The full table will be made available online.}
\label{tab:data} 
\begin{tabular}{l l l l l l l l l l l l l}
\hline
\hline
Bin Number &  RA &  DEC & x[``] &  y[``] & $v_{star}$ [km/s] & $dv_{star}$ [km/s] & $\sigma_{star} $ [km/s] & $d\sigma_{star}$ [km/s] & $h3_{star}$ & $dh3_{star}$ & $h4_{star}$ & $dh4_{star} $\\ 
  \hline
    0 & 00:42:44.23 & +41:16:13.98  &   2.65   &  4.57 & -298.65 & 1.88 & 150.80 &    2.21 &  0.00 & 0.01 & 0.01 & 0.01\\
    1 & 00:42:44.46 &+41:16:18.55  &   0.08   &  9.14 &-296.01 & 2.21 & 153.82 &     2.63 & 0.09 &0.01 & 0.02 & 0.01\\
    2 & 00:42:44.27 &+41:16:20.78  &    2.30  &  11.38 & -291.01 & 2.18 & 162.39 &    2.44 & -0.02 & 0.01 & 0.00 &0.01 \\
    3 & 00:42:43.99  &+41:16:18.52  &   5.38   & 9.11  &-292.93 &2.43 & 162.86 &    2.94  &0.02 &0.01 &  0.03 &0.01 \\
     4 & 00:42:44.73 &+41:16:20.57  &  -2.92   & 11.16 &-283.70 & 2.10 & 155.09 &    2.52 & -0.04 & 0.01 & 0.01 &0.01 \\
... \\
    \hline

\end{tabular}
 \end{table}
  \begin{table}
\caption{Kinematic data of the gas. The full table will be made available online.}
\label{tab:Kinematic_gas_data} 
\begin{tabular}{l l l l l l l l l}
\hline
\hline
Bin Number &  $v_{[OIII]}$ [km/s] & $dv_{[OIII]}$ [km/s] & $\sigma_{[OIII]} $ [km/s] & $d\sigma_{[OIII]}$ [km/s] & $v_{[OIII],2}$ [km/s] & $dv_{[OIII],2 }$ [km/s] & $\sigma_{[OIII],2 } $ [km/s] & $d\sigma_{[OIII],2 }$ [km/s]\\ 
  \hline
  \hline
0 &      -    &      -      &     -   &  - & -286.63 &  5.87 &   51.05 &      6.03 \\
1 &-322.91  &  8.02 &  62.89 &    7.43 & -231.25 &    6.53 &    15.61 &      8.95  \\
2 &-327.63  &  1.56 &   47.23 &   1.60 &           -   &       -     &       -     &        -  \\
3 &-364.28  &  2.06 &  29.62 &    2.19 & -266.52 &    4.20 &    27.97 &       4.47  \\ 
4 & -140.87  &  7.29 &  36.35 &    7.58 & -302.31 &   1.63 &    34.70 &     1.69 \\ 
...\\
\hline
 \end{tabular}
 \end{table}
 
\begin{table} 
 \caption{Photometric data of the gas. The full table will be made available online.}
\label{tab:Photometric_gas_data} 
\begin{tabular}{l l l l l l l l l l l l l}
\hline
\hline
Bin Number & $f_{H\beta}$ & $df_{H\beta}$& $f_{[OIII]}$ & $df_{[OIII]} $ & $ f_{[NI]}$ & $df_{[NI]}$ &$f_{H\beta,2}$ & $df_{H\beta,2}$ &  $f_{[OIII],2}$ &  $df_{[OIII],2} $ & $f_{[NI],2}$ &$df_{[NI],2}$\\
Flux units: $10^{15}$ [erg/s/cm$^2$] &  &    &                 &            &              &             &               &                 &                 &                   &              &              \\
  \hline
  \hline
    0  &     -       &          -        &         -      &           -       &          -      &           - & 2.3  &  0.4 & 2.8 & 0.4 & 1.1 &  0.3 \\
    1  &   2.7       & 0.4   & 5.3 & 0.7 & 7.4 & 0.2 & 0.1 & 0.1 & 0.5 & 0.3 &   0.0      &         0.0 \\
    2  & 3.2 & 0.2 & 7.1 & 0.3 & 1.4 & 0.2 &    -  &                -    &           -           &      -              &   -             &    - \\
    3  & 1.5 &  0.2 & 3.8 & 0.3 & 6.6 & 0.2 & 1.1 & 0.2 & 1.8 & 0.3 & 0.3 & 0.2 \\
    4  & 5.4 & 0.2 & 1.1 & 0.3 & 0.3 & 0.2 & 1.8 & 0.2 & 4.4 & 0.3 & 0.7 & 0.2 \\
    ...\\
    \hline
 \end{tabular}
 \end{table}
\end{landscape}

\end{document}